\documentclass{./template/articlex}

\begin{document}


\title{\textsc{The Hidden Subsidy of the Affordable Care Act}\thanks{We thank Tracy Miller, Brian Blase, and Drew Gonshorowski for helpful comments.}\\
	$~$\\}

\medskip

\author{\textbf{Liam Sigaud \protect\thanks{Mercatus Center at George Mason University. Email: lsigaud@mercatus.gmu.edu.}} \\  
\and
\textbf{Markus Bjoerkheim \protect\thanks{Mercatus Center at George Mason University. Email: mbjoerkheim@mercatus.gmu.edu.}} \\ 
\and
\textbf{Vitor Melo \protect\thanks{Clemson University and Knee Regulatory Research Center. Email: vmelo@clemson.edu.}} \\ 
}

\date{\today}              


\renewcommand{\thefootnote}{\fnsymbol{footnote}}

\singlespacing

\maketitle

\begin{abstract}
\noindent Under the ACA, the federal government paid a substantially larger share of medical costs of newly eligible Medicaid enrollees than previously eligible ones. States could save up to 100\% of their per-enrollee costs by reclassifying original enrollees into the newly eligible group. We examine whether this fiscal incentive changed states' enrollment practices. We find that Medicaid expansion caused large declines in the number of beneficiaries enrolled in the original Medicaid population, suggesting widespread reclassifications. In 2019 alone, this phenomenon affected 4.4 million Medicaid enrollees at a federal cost of \$8.3 billion. Our results imply that reclassifications inflated the federal cost of Medicaid expansion by 18.2\%.

\bigskip
			
			\noindent\emph{JEL Classification: H71, H77, I13, I18 } 
			
			\bigskip
			
			\noindent\emph{Keywords: Political Incentives, State Spending, Medicaid, Affordable Care Act} 
\end{abstract}

	\maketitle

\medskip

\thispagestyle{empty}

\clearpage

\onehalfspacing
\setcounter{footnote}{0}
\renewcommand{\thefootnote}{\arabic{footnote}}
\setcounter{page}{1}

\doublespacing
\setcounter{footnote}{0}
\renewcommand{\thefootnote}{\arabic{footnote}}
\setcounter{page}{1}

\doublespacing

\section{Introduction} \label{Introduction}
Medicaid is the single largest source of health insurance in the United States. It provides coverage to an estimated 85 million people and costs the federal government and states approximately \$750 billion per year \citep{mitchell2023medicaid}. The Affordable Care Act (ACA) of 2010 made substantial changes to the program, including permitting states to expand Medicaid eligibility to all non-elderly adults with incomes up to 138 percent of the federal poverty level (hereafter, the ``new adult group"). States that opted to expand Medicaid received enhanced federal matching funds for the new adult group. From 2014 to 2019, states covered a median of about 40\% of the medical costs for the original Medicaid population,\footnote{Throughout this paper, we use the term ``original Medicaid population" or ``original Medicaid enrollees" to refer to those, such as poor children and people with disabilities, who were eligible for Medicaid under pre-ACA eligibility rules.} but at most 7\% of the medical costs of enrollees in the new adult group. These provisions implied that the median state could save at least 82.5\% of their per-enrollee costs by reclassifying members of the original Medicaid population into the new adult group. We define ``reclassification" as enrollment in the new adult group when, in the absence of the ACA, the individual would been a part of the original Medicaid population.\footnote{This definition accommodates several forms of reclassification, ranging from deliberate, improper actions by state Medicaid administrators to legitimate shifts in enrollment stemming from natural life cycle events. These possibilities are discussed in more detail in Sections \ref{Reclassifications} and \ref{results}.} Given that Medicaid expenditures represent about one-fifth of states' general fund expenditures on average \citep{macpac_state_medicaid_funds}, reclassifications could represent a substantial hidden subsidy from the federal government to states.

We examine the existence and extent of the ACA's hidden subsidy by investigating the effect of expansion on enrollment in the original Medicaid population and estimating the fiscal impact of these enrollment changes. To quantify possible reclassifications, we examine the change in the original Medicaid population in expansion states relative to non-expansion states. Drawing on state administrative records from 2014 to 2019 and leveraging variation in the implementation of Medicaid expansion across states and time, we find that Medicaid expansion is associated with an average decline of 9.93\% in the number of original Medicaid enrollees. In 2019 alone, this represents 4.4 million fewer original Medicaid enrollees.

Since a Medicaid beneficiary's reclassification is primarily an administrative matter concerning states' requests for federal reimbursement, Medicaid enrollees would likely have no knowledge of how they were being classified. Nor would reclassifications necessarily affect enrollees' coverage or benefits. In some cases, such reclassifications were permitted under the ACA and subsequent federal rulemaking.\footnote{We discuss these dynamics in more detail in Section \ref{Reclassifications}.} In other cases, states may have deliberately reclassified enrollees in violation of federal law. States may also have incorrectly classified people into the new adult group due to carelessness or poor training of case managers. Irrespective of their legal status, such reclassifications are financially attractive to states.

Reclassifications, and the associated hidden subsidy, had substantial fiscal implications for states and the federal government. Our results indicate that the fiscal impact of Medicaid expansion on the U.S. Treasury would have been substantially lower if the effect we document had not occurred. Our estimates imply that the federal government distributed \$52.9 billion to states from 2014 to 2019 as a result of these reclassifications, including \$8.3 billion in 2019 alone. Based on these results, we revise CBO's estimates of the federal fiscal impact of Medicaid expansion and find that reclassifications increased federal costs by 18.2\% \citep{cbo2019}.

While many factors influence Medicaid enrollment, we argue that rival explanations are inadequate to account for our results. Where possible, we subject alternative theories to empirical scrutiny (see Sections and \ref{main_results} and \ref{robustness}). For example, we account for possible state-level changes to income thresholds for Medicaid eligibility, as well as changes in other administrative practices in Medicaid. We discuss other competing explanations --- including other changes to the health care system introduced by the ACA, such as the availability of premium tax credits for certain low- and middle-income households --- in greater detail in Section \ref{other_factors}. In our view, none provide a convincing explanation for the enrollment patterns that unfolded from 2014 to 2019. We conclude, therefore, that reclassifications played a key role in shrinking the size of the original Medicaid population in expansion states, relative to what it would have been absent the reform.

Previous research based on household surveys show that Medicaid expansion triggered a robust ``woodwork effect" --- i.e., large increases in enrollment among already-eligible people ``coming out of the woodwork" \citep{frean2017premium,hudson2017medicaid,gruber2019affordable, sacarny2022out}. This increase in awareness of Medicaid eligibility and enrollment of already-eligible people suggests that expanding Medicaid would increase the original population relative to a counterfactual where a state does not expand. However, our analysis of state administrative records indicates the opposite effect, suggesting that states engaged in large-scale reclassifications from the original Medicaid population to the new adult group that swamped the magnitude of the ``woodwork effect." Thus, our results should be considered a lower-bound estimate of reclassifications.

In addition to exposing a previously overlooked fiscal effect of the ACA, our work sheds light on several facts related to Medicaid expansion that have not been well understood: 1) Despite larger-than-expected enrollment, the fiscal impact of Medicaid expansion on states has been small \citep{sommers2017federal, gruber2020fiscal}. In fact, for some states, Medicaid expansion appears to have been a net fiscal benefit \citep{levy2020macroeconomic,simpson2020implications}. The hidden subsidy we document helped states offset the direct costs of Medicaid expansion; 2) Projections of the size of the new adult group have been greatly exceeded \citep{blase2016}. Reclassifications from the original Medicaid population to the new adult group, which was not contemplated by forecasters, may be an important mechanism behind these discrepancies; and 3) Per enrollee spending on the new adult group has been substantially higher than expected, and the ratio of per enrollee spending in the new adult group to other non-elderly adults on Medicaid has also exceeded actuarial expectations \citep{cms2013,cms2018}. Our results are consistent with this pattern. Members of the original Medicaid group tend to have higher medical spending than members of the new adult group, so reclassifications from the former to the latter would tend to increase per enrollee spending in the new adult group.


\section{Background and Policy Context} \label{Policy_Context}

\subsection{The Affordable Care Act and Medicaid Financing}

The expansion of Medicaid, arguably the centerpiece of the ACA's coverage provisions, has been responsible for a substantial decline in the uninsured rate among working-age Americans \citep{butler2016future}. The implications of this expansion for the health system and population health have been extensively studied \citep{buchmueller2016providers,peng2017does,huh2021medicaid,zhang2021does,neprash2021effect,miller2021medicaid,nikpay2022medicaid}, adding to other work on the effects of eligibility changes in public health insurance programs \citep{de2012effect,arenberg2024impact}. To date, forty states and the District of Columbia have expanded Medicaid under the ACA. In 2019, 12 million members of the new adult group were enrolled in Medicaid, accounting for about 16.2\% of total enrollment in the program.

Less is known about the impact of reforms made under the ACA to Medicaid's financing structure. Although Medicaid is operated by the states, the federal government contributes the majority of the program’s funding. Contributions by states accounted for approximately 31 percent of total Medicaid spending in 2021; the federal government paid 69 percent. Still, Medicaid represents a large and growing share of state budgets. In 2016, Medicaid accounted for nearly 20 percent of states' general fund expenditures, roughly double the program's share in the early 1990s \citep{macpac_state_medicaid_funds}. Medicaid is, by a wide margin, the most prominent example of fiscal federalism in the U.S.

Most federal Medicaid dollars are distributed to states on the basis of a formula that provides more assistance to states with low per capita personal income relative to the national average. The Federal Medical Assistance Percentage (FMAP), the share of Medicaid benefit spending reimbursed by the federal government, generally ranges from 50 percent (the statutory minimum) to about 77 percent, depending on the state. Over the decades, however, federal rules governing Medicaid funding have grown complex, with special treatment given to certain groups and service categories. The largest deviation from the traditional FMAP structure relates to Medicaid's expansion under the ACA to cover all low-income, non-elderly adults. Expansion states receive an FMAP rate for these enrollees that substantially exceeds the FMAP rate for most other Medicaid-eligible populations.\footnote{Other exceptions to the traditional FMAP rate include, e.g., enhanced federal matching for family planning services, smoking cessation programs for pregnant women, certain immunizations, and certain women with breast or cervical cancer. These carve-outs represent a very small proportion of total Medicaid spending, partly because the eligible populations are narrowly defined and partly because the FMAP rate enhancement is typically small. Therefore, we ignore these nuances for the purposes of our analysis.}

From 2014 to 2016, the federal government paid 100\% of the medical costs of the new adult group, declining to 95\% in 2017, 94\% in 2018, 93\% in 2019, 90\% in 2020, and remaining at 90\% in perpetuity. These enhanced federal reimbursement rates, which, unlike the FMAP rates for most of the rest of the Medicaid population, are not dependent on state average income, were designed to ease the fiscal burden on states and increase political support for the law. 

In 2014, when the new adult group was reimbursed by the federal government at a rate of 100\%, the FMAP rate for the original Medicaid population ranged from 50\% to 73\%, depending on the state; 27 states received FMAP rates below 60\%. In 2019, despite the FMAP rate for the new adult group declining to 93\%, the gap with the original population's FMAP rate remained large. That year, FMAP rates ranged from 50\% to 76\%, with 26 states received FMAP rates below 60\%. Moreover, since states that opted to expand Medicaid tend to have higher average incomes than non-expansion states, this group has disproportionately low FMAP rates for its original population, resulting in an even larger spread between the FMAP rate for the new adult group and the FMAP rate for the original Medicaid population.

To put the difference in federal support between the new adult group and the original Medicaid population in perspective, the Coronavirus Aid, Relief, and Economic Security (CARES) Act of 2020, which provided additional Medicaid resources to states during the COVID-19 pandemic, increased the traditional FMAP rate by a mere 6.2 percentage points --- roughly one-fifth the size of the FMAP rate spread established by the ACA.

Previous research has shown that states are responsive to changes in federal Medicaid funding \citep{grannemann1983controlling}.  \citet{adams2001fiscal} find that states succeed in substituting federal funds for state revenues, resulting in a reduction in state tax burdens for Medicaid. \citet{leung2022state} exploits a kink in the match rate formula to estimate that a percentage point increase in the federal Medicaid match raises per-enrollee spending by 3\% to 6\%. \citet{bundorf2022responsiveness} estimate that the ACA's enhanced FMAP rate led states to increase spending per original Medicaid enrollee by approximately 15\%, showing that state Medicaid spending is sensitive to the magnitude of the federal subsidy. We extend this work by examining how states reacted to the unprecedented fiscal incentives to reclassify enrollees embedded within the ACA's Medicaid expansion.

 
\subsection{The Woodwork Effect} \label{woodwork_effect}

Economists have long recognized and sought to document the spillover effects of reforms to social assistance programs \citep{bartik2002spillover,baicker2005spillover,grabowski2006cost,mcinerney2017effects,carey2020impact}. Expanding public programs to cover a new group of people tends to increase enrollment among those who were already eligible under the pre-expansion eligibility criteria. This phenomenon --- known as the ``woodwork” or ``welcome mat" effect --- may be particularly strong when a program’s expansion is widely publicized. Millions of Americans are eligible for Medicaid but are not enrolled in the program \citep{sommers2011states}. Although foregoing Medicaid coverage may be a deliberate choice for some individuals, administrative barriers and lack of awareness of program rules may play a decisive role in many cases. The passage of the ACA, of which the expansion of Medicaid was a core component, generated substantial media coverage and considerable public interest. Many states advertised Medicaid expansion on billboards and in TV and radio ads, urging the public to check their eligibility \citep{artiga2013}. Moreover, the ACA instituted other policy changes, such as tax incentives to obtain health insurance and measures to streamline the Medicaid application process, that likely contributed to the woodwork effect.

The most reliable evidence of woodwork effects in Medicaid predates the ACA. \citet{sonier2013medicaid} estimates that health reforms adopted in Massachusetts in 2006, which align closely to key design features of the ACA, provoked large woodwork effects that substantially increased Medicaid enrollment. \citet{sacarny2022out} leverage data from the Oregon Health Insurance Experiment, in which Medicaid eligibility was determined by lottery, and calculate a short-run 6\% increase in child enrollments when adults in their household gained access to Medicaid. 

Several studies have sought to quantify the woodwork effect in the context of the ACA using household survey data, typically drawn from the American Community Survey, an annual, large-scale survey of U.S. households. \citet{frean2017premium} estimate that fully half of the impact on coverage attributable to Medicaid expansion in 2014-2015 came from the woodwork effect. The woodwork effect was found to be large in both expansion and non-expansion states. \citet{hudson2017medicaid} find that the ACA induced a large woodwork effect among children. They estimate that 710,000 low-income children gained Medicaid coverage through the woodwork effect in 2014-2015. \citet{mcinerney2021welcome} present evidence of a woodwork effect among seniors dually eligible for Medicare and Medicaid. They estimate the ACA increased Medicaid enrollment in this population by 4.4\%.

Since care provided to the original Medicaid population was reimbursed by the federal government at an FMAP rate of only 50\% to 77\% between 2014 and 2019,\footnote{The statutory maximum traditional FMAP rate is 83\%, but in practice no state's traditional FMAP rate exceeded 77\% during the 2014-2019 period.} the size of the woodwork effect is -- at least theoretically -- a key parameter in estimating the impact of Medicaid expansion on states' budgets. In certain states, these costs were expected to be non-trivial \citep{price2013economic}. 

However, our analysis of state administrative records suggests that previous research on Medicaid expansion and the woodwork effect, while perhaps an accurate reflection of the gains in Medicaid coverage among previously eligible individuals, should not be used to calculate the state fiscal costs of Medicaid expansion. This discrepancy arises because of trends in how states classified Medicaid enrollees when seeking federal reimbursement for program expenses. We find that a large number of individuals who otherwise would likely have been reported as belonging to the original Medicaid population were reclassified into the new adult group in expansion states.

\subsection{Reclassification of Enrollees to the New Adult Group under the ACA} \label{Reclassifications}

The ACA created several channels for states to shift Medicaid enrollees who would otherwise have be classified in the original Medicaid population (and been reimbursed at the traditional FMAP rate) to the new adult group under the enhanced FMAP rate. 

First, adult Medicaid enrollees who were not eligible for full benefits prior to the ACA's passage (e.g., individuals receiving family planning services under waivers granted by CMS or individuals eligible under special Medicaid rules for the ``medically needy") could be transferred to the new adult group under the ACA, and states could receive the enhanced FMAP rate for care provided to these individuals.

Second, the ACA created opportunities for individuals to join the new adult group prior to experiencing a health event (e.g., pregnancy or a disabling injury) that would otherwise have made them eligible for the original Medicaid population. For example, a woman may qualify for the new adult group and enroll in Medicaid before becoming pregnant. During her pregnancy, states are allowed to maintain her classification in the new adult group and receive enhanced FMAP rates for her pregnancy-related care. Similarly, enrollees in the new adult group who become disabled may remain in the new adult group. In the counterfactual where Medicaid expansion had not occurred, many pregnant women and people with disabilities would presumably have joined the original Medicaid population. Instead, Medicaid expansion ``siphons off" some of these enrollees, resulting in lower enrollment in the original Medicaid population and larger federal subsidies to states.

In addition to the mechanisms described above, which were authorized under the ACA, it is possible that states may have -- knowingly or unknowingly -- reclassified enrollees in violation of federal laws and regulations. Medicaid administrative tasks, including eligibility verification, data management, and reporting to CMS, are almost entirely controlled by states with minimal federal oversight. Moreover, CMS exerts little meaningful pressure on states to correct errors in eligibility classifications or deficiencies in data management practices. A recent report by the Government Accountability Office noted: ``While CMS is generally required to disallow, or recoup, federal funds from states for eligibility-related improper payments…, it has not done so for decades… [I]n July 2017, CMS issued revised procedures through which it can recoup funds for eligibility errors, beginning in fiscal year 2022” \citep{gao2020}. Consequently, during our entire post-treatment period (2014-2019), states faced no financial sanctions for eligibility errors.

The failure to properly determine enrollees' eligibility is widespread in Medicaid \citep{blase2022}. Audits of state Medicaid records carried out by the U.S. Department of Health and Human Services provide direct evidence that misclassifications of Medicaid enrollees -- including individuals who should be classified in the original Medicaid population being reported as belonging to the new adult group -- occur on a fairly large scale. Investigations conducted in 2018 and 2019 in California, New York, and Colorado (all of which expanded Medicaid in 2014) suggest that as many as 28.3\% of individuals classified as new adult group enrollees may be ineligible \citep{oig2018,oig2019,oig2019ny}, a figure that matches closely with our estimates. Based on the Payment Error Rate Measurement (PERM) system, CMS estimated in 2019 that improper eligibility determinations accounted for 8\% of federal Medicaid payments, amounting to approximately \$32.3 billion \citep{cms2019perm}.

\section{Data} \label{Data}

We construct a balanced state-level panel of the original Medicaid population from 2006 to 2019. We exclude later years because the COVID-19 pandemic, and the government response to the public health crisis, substantially affected Medicaid enrollment and altered states' fiscal incentives. Most importantly, states paused their normal eligibility redetermination processes from early 2020 to early 2023, leading to a nationwide surge in Medicaid enrollment. Other temporary policies included a 6.2 percentage point increase in states' traditional FMAP rates, which narrowed the FMAP spread between the traditional FMAP rate and the enhanced FMAP rate. It would be difficult to disentangle the enrollment effects of Medicaid expansion from the effects of these forces. Moreover, we believe the future of Medicaid is more likely to resemble the 2014-2019 period than the anomalous pandemic years, so focusing on the pre-pandemic period is likely to yield more valuable insights.

Our data comes from two sources. We obtain data for the years 2006-13 from issue briefs published by the Kaiser Family Foundation. These data were compiled by Health Management Associates, a research and consulting firm, based on internal state Medicaid enrollment records.\footnote{CMS does not publicly release state-level Medicaid enrollment data for the years 2006-2013. The figures reported represent ``point-in-time” monthly Medicaid enrollment counts for the month of June each year (enrollment for the month of December of each year was also reported but not used in our analysis). Every person with Medicaid coverage was counted as an enrollee with the exception of family planning waiver enrollees and pharmacy plus waiver enrollees. No adjustment was made for other persons who were enrolled in Medicaid categories with less than full coverage. Therefore the enrollment figures include a small number of individuals who are covered by Medicaid only for emergency services or services related to breast and cervical cancer, and persons with Medicare and Medicaid dual eligibility enrolled as Qualified Medicare Beneficiaries (QMBs), Specified Low-Income Medicare Beneficiaries (SLMBs) or Qualified Individuals (QIs) for whom Medicaid pays a portion of Medicare premiums, copays, and deductibles. Persons in state-only health coverage programs and Medicaid expansion CHIP enrollees not funded by Medicaid are excluded.} Our second source of data, which covers the 2014-19 period, is Medicaid enrollment reports submitted by states to the Centers for Medicare \& Medicaid Services (CMS) through the Medicaid Budget and Expenditure System (MBES).\footnote{Both sources capture only individuals whose coverage is funded through Medicaid (Title XIX of the Social Security Act); children and young adults funded through CHIP are excluded.} Post-ACA enrollment information is a count of unduplicated individuals enrolled in the state’s Medicaid program at any time during each month in the quarterly reporting period. The enrollment data identifies the total number of Medicaid enrollees and, for states that have expanded Medicaid, provides specific counts for the number of individuals enrolled in the new adult group.\footnote{The new adult group consists of two distinct populations: newly-eligible and non-newly-eligible. Non-newly-eligible enrollees are a small, special class of Medicaid recipients already enrolled in Medicaid when the ACA was passed. To calculate the number of enrollees in the original Medicaid population, we subtract the number of newly eligible enrollees from the total number of Medicaid enrollees. This will tend to bias our results against finding a decline in the size of the original Medicaid enrollment, since it is possible that states have reclassified enrollees from the original Medicaid population to the non-newly-eligible group \citep{bundorf2022responsiveness}.} Enrollment figures for the month of June were used for each year analyzed. We define our dependent variable as the natural log of the number of individuals in the original Medicaid population.

The use of two different datasets to track Medicaid enrollment across time is not ideal, since differences in how each dataset is collected and compiled could potentially influence our results. In our case, this concern is compounded by the fact that the endpoints of each dataset coincide with the beginning of treatment for the largest cohort of states. However, we know of no alternative source of publicly available yearly Medicaid enrollment figures at the state level. Moreover, in Section \ref{robustness} we perform several empirical tests to determine whether our approach affects the main results; we find no such evidence.

Many factors affect the size of the original Medicaid population. We explore a range of specifications with a range of state-level covariates that capture differences in Medicaid program rules, political conditions, demographics, and the state of the economy. We account for income eligibility thresholds for key subgroups within the original Medicaid population (children and parents), using data from the Kaiser Family Foundation. More stringent eligibility thresholds would tend to reduce the size of the original Medicaid population. Since Medicaid enrollment tends to be counter-cyclical, in some specifications we control for the state unemployment rate, the state poverty rate, the maximum level of welfare (TANF) benefits for a family of three, or the state food insecurity rate, all of which measure economic distress; these variables come from the University of Kentucky's Center for Poverty Research. We also consider the demographic composition of the state population (proportion non-White), since Medicaid enrollment varies across racial groups, as well as the size of the state population (both from the Census Bureau). Finally, in some specifications we use data from the University of Kentucky's Center for Poverty Research to adjust for the political party of the chief executive (governor for states and mayor for the District of Columbia) to account for potential differences in how the Medicaid program is administered. We control for baseline values of our state covariates in the last period before Medicaid expansion was implemented. The path of the original Medicaid population (in the absence of expansion) likely depends on these covariates, so a conditional parallel trends assumption may be more plausible than an unconditional parallel trends assumption.

We present descriptive statistics (means and standard deviations) of our outcome variable, as well as all state-level covariates, in Table \ref{summary_stats_table}. Expansion and control states are broadly similar across several dimensions, including population size, racial diversity, and economic performance. Unsurprisingly, expansion states are substantially more likely to have a Democratic governor, provide more generous TANF benefits, and have higher income limits for parents on Medicaid.

\bigskip\smallskip


\begin{table}[H]
\caption{\label{summary_stats_table}Descriptive Statistics}
\centering
\footnotesize
\begin{tabular}{@{} m{7cm}m{1.7cm}m{1.7cm}m{1.7cm}m{1.7cm}@{}}
\hline \hline
 & \multicolumn{2}{c}{Expansion States} & \multicolumn{2}{c}{Control States} \\ 
\\[-5.5ex]
\cmidrule(lr){2-3} \cmidrule(lr){4-5}
\\[-5ex]
\textit{Variable} & Mean & SD & Mean & SD \\
\\[-5ex]
 & (1) & (2) & (3) & (4) \\
 \\[-5ex]
\hline
Original Medicaid Population (ln) &       13.26&       1.119&       13.33&       1.114\\
\\[-5ex]
Eligibility Limit, Children (Proportion of FPL) &       2.563&       0.583&       2.231&       0.385\\
\\[-5ex]
Eligibility Limit, Parents (Proportion of FPL) &       1.167&       0.537&       0.542&       0.354\\
\\[-5ex]
State Unemployment Rate (\%)   &       5.888&       2.192&       5.543&       2.230\\
\\[-5ex]
Governor's Political Party (1 = Dem) &       0.569&       0.496&       0.189&       0.392\\
\\[-5ex]
State Population (ln) &       15.08&       1.055&       15.24&       0.980\\
\\[-5ex]
Non-White (Proportion of State) &       0.211&       0.142&       0.232&       0.128\\
\\[-5ex]
Maximum TANF Benefits (\$) &       485.5&       163.6&       347.9&       139.1\\
\\[-5ex]
Food Insecurity (Proportion of State) &       0.132&      0.0338&       0.150&      0.0329\\
\\[-5ex]
Poverty Rate &       12.46&      3.464&       13.62&      3.260\\
\\[-5ex]
\hline
\\[-5ex]
Observations        &         476 &            &         238&            \\\hline\hline
\\[-6.25ex]
\end{tabular}
\begin{tablenotes}
\footnotesize
    \item \textit{Notes}: This table presents descriptive statistics for variables used in our analysis, split by state Medicaid expansion status. Expansion states consist of 33 states (and the District of Columbia) that expanded Medicaid under the ACA before the end of 2019. Control states consist of 17 states that had not expanded Medicaid by the end of 2019. FPL refers to the Federal Poverty Level (in 2019, approximately \$25,750 for a family of four). \textit{Sources}: Medicaid original population is drawn from issue briefs published by the Kaiser Family Foundation and enrollment reports submitted by states to CMS. Unemployment rates are from the Bureau of Labor Statistics. Medicaid eligibility limits are from the Kaiser Family Foundation. State population and proportion of population non-White come from the Census Bureau. The political party of the governor (or mayor, in the case of the District of Columbia), poverty rates, TANF benefits, and food insecurity rates come from the University of Kentucky's Center for Poverty Research.
\end{tablenotes}
\end{table}


\section{Empirical Strategy} \label{Emperical_Strategy}
To identify the effect of Medicaid expansion on enrollment, we leverage variation in the adoption of Medicaid expansion across geographies and time, comparing trends between states that opted to expand Medicaid under the ACA and states that did not. Since non-expansion states did not experience a relative change in their FMAP rates to cover different groups of Medicaid recipients, they represent a natural control group to test our reclassification hypothesis. Historically, two-way fixed effects (TWFE) regressions have served as the workhorse models for estimating causal effects in the context of staggered policy adoption. However, recent studies have shown that the TWFE estimator can yield inconsistent and misleading estimates of the average treatment effect on the treated (ATT) in the presence of treatment effect heterogeneity between groups or across time \citep{de2020two,borusyak2021revisiting,callaway2021difference,goodman2021difference,imai2021use,sun2021estimating}. 

These concerns related to TWFE models apply to our setting, in which states expanded Medicaid at different times (the first expansions in our data occur in 2014 and the last occurs in 2019; see Table \ref{expansion_timing_table} in the Appendix for details on the treatment timing of individual states).\footnote{For this reason, the preliminary descriptive evidence presented in Figure \ref{figure1} and Table \ref{tab:simple_dd} in Section \ref{results} is based on simple comparisons of the initial expansion cohort of states (i.e., the 25 states (including the District of Columbia) that expanded Medicaid in January 2014) and the 17 states that did not expand Medicaid by the of 2019, when our sample ends. In these exhibits, we exclude the nine states that expanded Medicaid between February 2014 and December 2019.} To overcome these limitations, in our main results we implement the robust difference-in-differences estimator proposed by \citet{callaway2021difference}. This approach allows us to retain all states in our sample, including those that expanded Medicaid after the initial cohort in January 2014. The Callaway Sant'Anna method delivers consistent ATT estimates even in the presence of arbitrary heterogeneous treatment effects by shutting down problematic 2 $\times$ 2 difference-in-differences comparisons between newly treated and already treated states. We implement the augmented inverse-probability weighting estimator described in \citet{callaway2021difference}, in which both the treatment and outcome are modeled; recovering consistent estimates depends only on correctly specifying one of the models. 

For our comparison group, we use only states that did not expand Medicaid before the end of 2019, when our sample ends (i.e., ``never-treated" states). An alternative approach would be to include ``not yet" treated states in the comparison group. We choose to restrict the comparison group to ``never-treated" states for several reasons. First, our data includes a relatively large number of 17 ``never-treated" states and a relatively small number of late-expanding states (i.e., those that would serve as additional controls under the ``not yet" option). Second, ``never-treated" states are broadly similar to treated states, with geographic representation in the South, West, and Midwest. Third, the economic conditions during early and late treatments differ. Fourth, the parallel trends assumption is different between the two choices, and its interpretation is more straightforward when the comparison group is limited to ``never-treated" states \citep{callaway2021difference}.

There are a variety of ways to represent the results from the \citet{callaway2021difference} estimator. In our main results (Table \ref{tab:main_results}), we focus on the overall ATT, which is a (simple) weighted average of each $ATT(g,t)$, where $g$ denotes the treatment group and $t$ denotes the year. This calculation aggregates the ATTs within all treatment groups and time periods. In Figure \ref{main_results_plot} we also present dynamic ATTs across treatment event time. In addition to highlighting treatment effects with respect to length of exposure to treatment, this dynamic specification allows us to assess the plausibility of the parallel trends assumption.

\section{Results} \label{results}

\subsection{Main Results} \label{main_results}

\begin{figure}[H]
\centering
\caption{\label{figure1} Original Medicaid Population Enrollment (\% Change Compared to 2013)}
\includegraphics[width=0.75\textwidth]{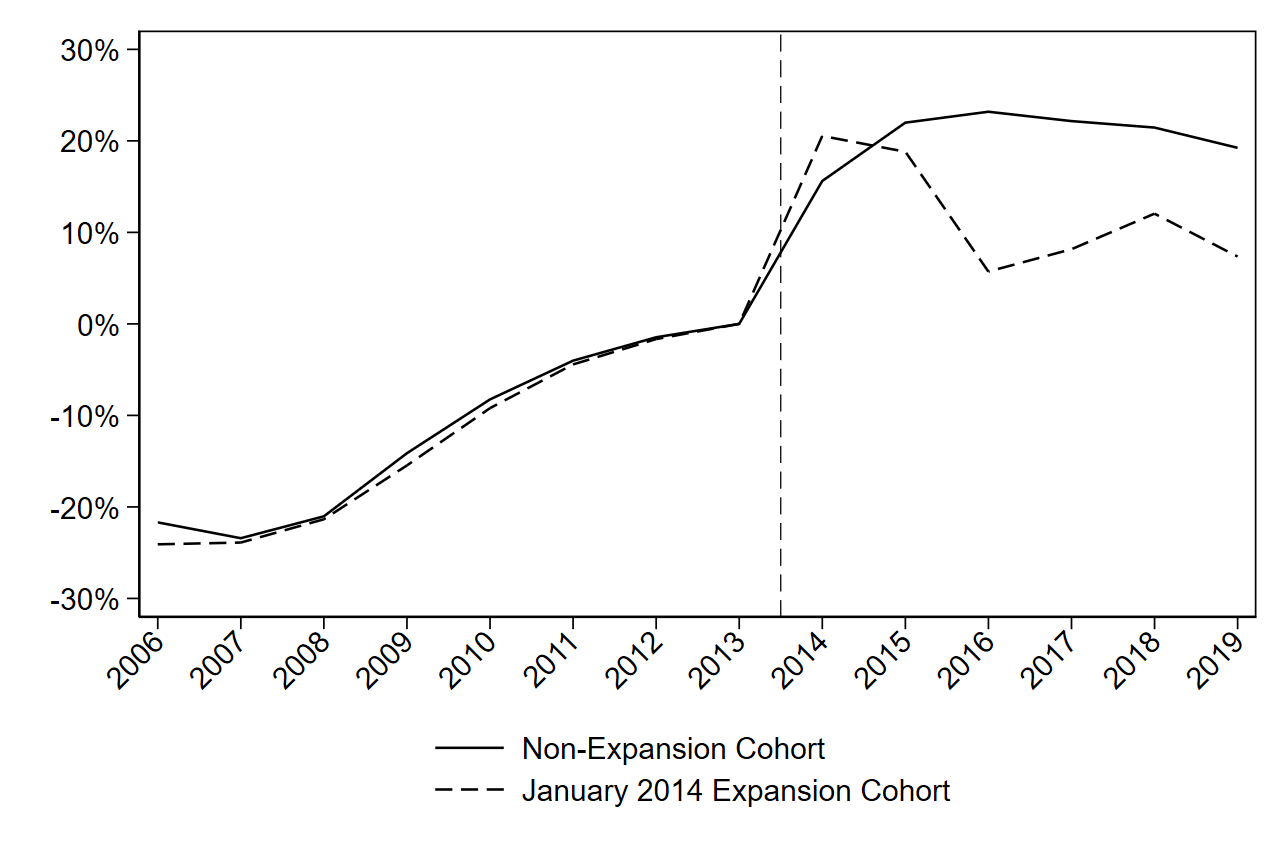}
\caption*{{\footnotesize 0.5\textit{Notes}: 
This figure plots the change in the size of the original Medicaid population in non-expansion states and in states that expanded Medicaid in January 2014. Values are normalized to zero in 2013, the last pre-expansion year. We use enrollment figures for the month of June in each year. The vertical dashed line denotes the implementation of the ACA's Medicaid expansion. The January 2014 expansion cohort consists of 25 states (including the District of Columbia). The non-expansion cohort consists of 17 states that have not expanded Medicaid under the ACA (as of April 2024) as well as states that expanded after 2019. The remaining nine states expanded Medicaid in a staggered fashion between February 2014 and December 2019; for simplicity, we omit these states from the graph. Data was compiled by the authors from Kaiser Family Foundation issue briefs (for the years 2006-2013) and reports from the Medicaid Budget and Expenditure System (for the years 2014-2019); see section \ref{Data} for more details. We define the original Medicaid population as total Medicaid enrollment minus the number of enrollees reported by states as ``newly eligible" under the ACA.}}
\end{figure}

In this section, we discuss our empirical results. Before turning to more sophisticated statistical methods, we present graphical evidence of longitudinal trends. Figure \ref{figure1} plots the change in the size of the original Medicaid population (measured in individuals enrolled in June of each year), contrasting states that expanded in January 2014 with those that had not expanded by the end of 2019. For each cohort and year, we sum enrollment across all states. For ease of comparison, for both cohorts we express the change in enrollment relative to 2013, the last pre-expansion year. From 2006 to 2013, both cohorts tracked closely together. For both groups, enrollment in 2006 was approximately 21-24\% lower than in 2013. From 2013 to 2014, the first treated year, both cohorts continued to follow very similar growth paths, with expansion states showing slightly larger gains in enrollment. Beginning in 2015, however, the cohorts began to diverge. Non-expansion states continued to experience positive enrollment growth in 2015 and 2016, before declining gradually through 2019 --- broadly consistent with how one would expect Medicaid enrollment to evolve given the strengthening state of the national economy during this period and the national reach of the ``woodwork effect" triggered by Medicaid expansion. Meanwhile, states that expanded Medicaid in January 2014 reported negative enrollment growth in 2015 and 2016, followed by a small rebound in 2017 and 2018 and renewed decline in 2019. In total, from 2013 to 2019, enrollment in the original Medicaid group declined by 1.7\% in states that expanded in January 2014. Over the same period, non-expansion states reported a 19.2\% increase in enrollment.

\vspace{1cm}

Next, using the same data, we formalize this comparison by deriving simple difference-in-differences estimates of Medicaid expansion's effect on enrollment in the original Medicaid population. Table \ref{tab:simple_dd} compares changes in the state-reported size of the original Medicaid population in the pre-treatment period (2006-2013) and the post-treatment period (2014-2019) between the cohort of states that expanded in January 2014 and states that had not expanded by the end of 2019. In the pre-treatment period, the mean level of enrollment in the original Medicaid population in expansion states was 1.10 million, while the mean in non-expansion states was 0.92 million. In the post-treatment period, the mean in expansion states grew to 1.35 million, while the mean in the non-expansion cohort increased to 1.23 million. Hence, our simple difference-in-differences calculation implies that, on average, the original Medicaid population would have been larger by nearly 58,000 enrollees (or 4.29\%) in expansion states in the absence of the expansion.

\vspace{1cm}

\begin{table}[H]
\centering
\renewcommand{\arraystretch}{1.5}
\caption{Simple Difference-in-Differences Estimate}
\label{tab:simple_dd}
\begin{threeparttable}
\footnotesize
\begin{tabular}{l*{4}{c}}
\toprule \toprule
\\[-9ex]
 & Pre-treatment & Post-treatment & Difference (Pre/Post) & Difference-in-Differences \\
\midrule
\\[-9ex]
\emph{\underline{State Cohort}} & & & & \\
\\[-8ex]
Jan. 2014 expansion states & 1,101,718 & 1,350,646 & +248,928 & -57,894 \\ 
\\[-10ex]
Non-expansion states & 923,996 & 1,230,818 & +306,822 & \\
\bottomrule \toprule\\[-8.75ex]
\end{tabular}
\begin{tablenotes}
    \item \textit{Notes}: This table compares the average level of enrollment in the original Medicaid population between pre- and post-treatment periods and expansion and non-expansion states. We use enrollment figures for the month of June in each year. The January 2014 expansion cohort consists of 25 states (including the District of Columbia). The non-expansion cohort consists of 17 states that have not expanded Medicaid under the ACA (as of April 2024) as well as states that expanded after 2019. The remaining nine states expanded Medicaid in a staggered fashion between February 2014 and December 2019; for simplicity, we omit these states from our calculations. Data was compiled by the authors from Kaiser Family Foundation issue briefs (for the years 2006-2013) and reports from the Medicaid Budget and Expenditure System (for the years 2014-2019); see section \ref{Data} for more details. We define the original Medicaid population as total Medicaid enrollment minus the number of enrollees reported by states as ``newly eligible" under the ACA.
\end{tablenotes}
\end{threeparttable}
\end{table}

\begin{figure}[H]
\centering
\caption{\label{main_results_plot} Dynamic Effects of Medicaid Expansion on Enrollment}
\includegraphics[width=\textwidth]{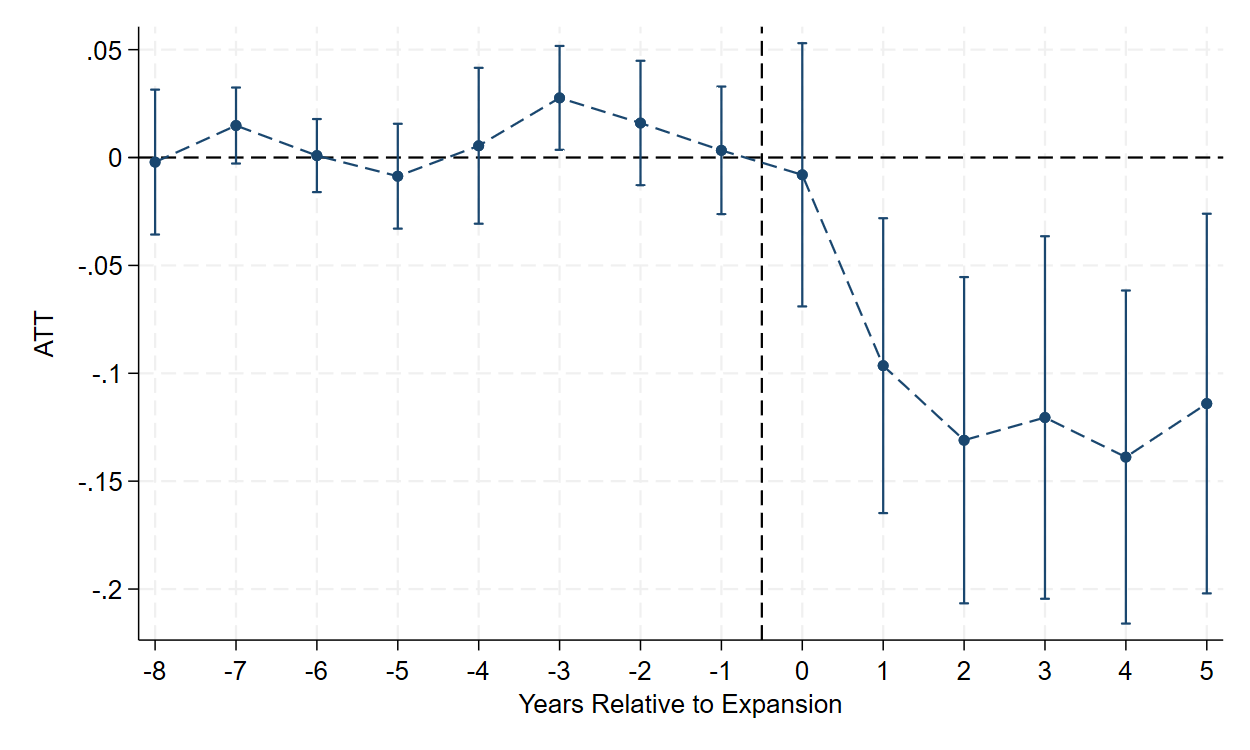}
\caption*{{\footnotesize \textit{Notes}: This plot shows dynamic effects across event time, based on our preferred specification (column (3) in Table \ref{tab:main_results}). Bars represent 95\% confidence intervals. The vertical dashed line represents the implementation of Medicaid expansion. We use enrollment figures for the month of June in each year. Data was compiled by the authors from Kaiser Family Foundation issue briefs (for the years 2006-2013) and reports from the Medicaid Budget and Expenditure System (for the years 2014-2019); see section \ref{Data} for more details. We define the original Medicaid population as total Medicaid enrollment minus the number of enrollees reported by states as ``newly eligible" under the ACA.} }
\end{figure}

While informative, the comparisons presented in Table \ref{tab:simple_dd} have three important shortcomings. First, they ignore potentially confounding factors. Second, they omit late-expanding states (i.e., those that expanded between February 2014 and December 2019). Third, they conceal the dynamic effects of Medicaid expansion across different treatment periods. In Table \ref{tab:main_results} and Figure \ref{main_results_plot}, we address each of these limitations by implementing the difference-in-differences estimator described in \citet{callaway2021difference}. Since our dependent variable is the log of enrollment in the original Medicaid population, our regression coefficients can be interpreted as (approximate) percent changes. To obtain a baseline, in column (1) of Table \ref{tab:main_results} we drop late-expanding states and estimate the model without controls. The coefficient does not attain statistical significance (\textit{p} = 0.11) but is similar in magnitude to our implied estimate in Table \ref{tab:simple_dd}. Each of the other specifications presented in Table \ref{tab:main_results} include all states and account for the staggered adoption of Medicaid expansion across time. Column (2) shows the no-controls specification with all states. Once again, the coefficient is similar to the implied estimate from Table \ref{tab:simple_dd} but is not statistically significant (\textit{p} = 0.13). In column (3), we present our preferred specification, adding controls for the Medicaid income eligibility threshold for parents, the political party of the governor, and the state unemployment rate. These variables account for a range of possible sources of confounding. The Medicaid income eligibility threshold for parents reflects changes to eligibility affecting the original Medicaid population. We also adjust for the political party of the chief executive because Democratic and Republican governors may administer their Medicaid programs differently, in ways that are difficult to explicitly capture (e.g., the level of outreach to eligible populations). Finally, the state unemployment rate helps to isolate our estimates from the impact of economic shocks on Medicaid enrollment. The magnitude of the coefficient in column (3) is large and statistically significant at the 1\% level, indicating that Medicaid expansion leads to a 9.93\% decline in the size of the original Medicaid population. 

In columns (4) to (9) we present a range of alternative specifications using our preferred specification, column (3), as a baseline. Column (4) adjusts for the Medicaid income eligibility threshold among children (rather than parents), another major sub-group of the original Medicaid population. In column (5), we use the poverty rate as a proxy for state economic conditions, rather than the unemployment rate. Column (6) adds the log of state population to adjust for inter-state shifts in population. Column (7) adds the proportion of the state population that is non-white. In Column (8), to account for the fact that the generosity of safety-net programs may have spillover effects on enrollment in other programs \citep{schmidt2019impact}, we add the maximum level of TANF benefits for a family of three. Finally, column (9) shows the effect of using the food insecurity rate rather than the unemployment rate to measure economic distress. All alternative specifications yield similar results.

As a basis for later computations, we use the coefficient given in column (3), which is approximately in the middle range of our estimates.

\vspace{1cm}

\begin{table}[H]
\centering
\caption{Effects of Medicaid Expansion on Enrollment in the Original Medicaid Population}
\label{tab:main_results}
\begin{threeparttable}
\scriptsize
\begin{tabular}{@{} p{2.65cm}C{0.9cm}C{0.9cm}C{1.25cm}C{1.1cm}C{1.25cm}C{1.1cm}C{1.1cm}C{1.cm}C{1cm}@{}}
\toprule
 & (1) & (2) & (3) & (4) & (5) & (6) & (7) & (8) & (9) \\
\midrule
ATT & -0.0595 & -0.0505 & -0.0993\textsuperscript{***} & -0.0675\textsuperscript{**} & -0.1064\textsuperscript{***} & -0.0912\textsuperscript{**} & -0.1021\textsuperscript{**} & -0.0638\textsuperscript{*} & -0.0895\textsuperscript{*} \\
 & (0.0370) & (0.0336) & (0.0348) & (0.0320) & (0.0369) & (0.0382) & (0.0428) & (0.0331) & (0.0348) \\
\addlinespace
Gov's Political Party & & & \checkmark & \checkmark & \checkmark & \checkmark & \checkmark & \checkmark & \checkmark \\
Elig. Limit, Parents & & & \checkmark & & \checkmark & \checkmark & \checkmark & \checkmark & \checkmark \\
Unemployment Rate & & & \checkmark & \checkmark & & \checkmark & \checkmark & \checkmark & \\
Elig. Limit, Children & & & & \checkmark & & & & & \\
Poverty Rate & & & & & \checkmark & & & & \\
ln(State Population) & & & & & & \checkmark & & & \\
Non-White (\% of State) & & & & & & & \checkmark & & \\
TANF Benefits & & & & & & & & \checkmark & \\
Food Insecurity Rate & & & & & & & & & \checkmark \\
\addlinespace
N & 602 & 714 & 714 & 714 & 714 & 714 & 714 & 714 & 714 \\
\bottomrule
& & & & & & & & \\[-6.25ex]
\end{tabular}
\begin{tablenotes}
\footnotesize 
    \item \textit{Notes}: This table shows estimates of the ATT of Medicaid expansion on the size of the original Medicaid population (logged) across a range of models, all of which use the staggered difference-in-differences estimator described by \citet{callaway2021difference}. The comparison group is "never" treated units. The specification in column (1) is without controls and includes only states that expanded in January 2014. All other models include all states. Standard errors (clustered by state) are reported in parentheses. * $p<0.10$, ** $p<0.05$, *** $p<0.01$.
\end{tablenotes}
\end{threeparttable}
\end{table}

Figure \ref{main_results_plot} plots the dynamic treatment effects derived from our preferred specification (column (3) in Table \ref{tab:main_results}). The event study is generally supportive of the parallel trends assumption, showing little evidence of differential trends in the periods preceding the expansion of Medicaid. We also note the magnitude of the effect, with the exception of the first year of Medicaid expansion's implementation (Year 0 in Figure \ref{main_results_plot}), is statistically significant and roughly constant throughout the post-expansion period.

\vspace{1cm}

We offer two possible explanations for the absence of a clear effect in the first treated year (Year 0 in Figure \ref{main_results_plot}), which for the vast majority of states corresponds to 2014. First, there is evidence that the woodwork effect in expansion states was particularly strong in 2014 \citep{ian2014,frean2017premium}, which would have counteracted, and perhaps fully offset, the effects of reclassifications on the number of people in the original Medicaid population. The woodwork effect likely waned in later years, partly due to the mechanical decline in the number of eligible-but-not-enrolled individuals and partly because by mid-2015, some states had already shifted away from large-scale community outreach efforts to raise awareness about Medicaid \citep{artiga2015}. Second, it may have taken time for state policymakers and administrators to adapt to the new bureaucratic structures established by the ACA and scale-up reclassification efforts.

A range of additional checks are shown in the Appendix. Table \ref{tab:weighted} gives results weighted by each state's 2013 Medicaid population to ensure that our findings are not unduly influenced by small states; dynamics effects of the main specification, with weights, are shown in Figure \ref{appendix_results_plot_weighted}. Weighting yields similar, or slightly larger, effects. In Table \ref{tab:robustnesspreferred} we show the sensitivity of our main specification to different choices in defining the treatment group. Specifically, we show the effects of dropping late-expanding states (i.e., those that expanded Medicaid after January 2014), states that implemented early ACA expansions during 2010-2012, and states that covered low-income childless adults prior to the ACA's passage in 2010. We report coefficients from both weighted and unweighted models. All specifications remain statistically significant at the 5\% or 1\% level, and seven out of nine alternative samples yield treatment effects larger than our main estimate. Finally, in Table \ref{tab:early_expanders}, we limit the sample to states that expanded in January 2014 and present the same set of specifications as in Table \ref{tab:main_results}; dynamics effects of the main specification with this narrower sample are shown in Figure \ref{appendix_2014_states_results_plot}. Results are consistent.

\subsection{Other ACA-Related Factors} \label{other_factors}

Before turning to the fiscal implications of our empirical results, we consider several alternative explanations to our reclassification hypothesis and argue that no other explanation can plausibly account for the large decline in the original Medicaid population in expansion states relative to non-expansion states over the 2014-2019 period. In the discussion that follows, we focus on major provisions of the ACA (other than Medicaid expansion) that had substantial effects on the U.S. health care system.

\vspace{0.5cm}
\noindent \textit{Premium Tax Credits} --- The ACA created a system of tax subsidies (in the form of premium tax credits, or PTCs) to help lower- and moderate-income Americans purchase private health insurance on the non-group market. During our sample period, households in expansion states were eligible for PTCs if their income fell between 138\% and 400\% of the FPL. In non-expansion states, the households earning between 100\% and 400\% of the FPL were eligible for PTCs. The lowest-income households received more generous subsidies, with premium contributions capped at 2\% of their annual income. For some original Medicaid enrollees, transitioning to private coverage --- possibly perceived as being higher quality than Medicaid --- may have been appealing. Yet there is little reason to think that such transitions are driving our results. First, under federal law individuals eligible for Medicaid are not eligible for PTCs, so this hypothesis requires millions of households to have strategically adjusted their income or other characteristics to gain PTC eligibility. Second, this hypothesis requires PTC-induced transitions from Medicaid to ACA plans to have been substantially larger in expansion states than in non-expansion
states. Yet despite more than 8 million Americans signing up for ACA plans during the 2013-2014 open enrollment period \citep{aspe_brief2014}, we see no differential effect on original Medicaid enrollment that year (see Figure \ref{appendix_2014_states_results_plot}). Third, we note that the individual mandate was eliminated in 2019; yet we detect strong effects that year.

\vspace{0.5cm}
\noindent \textit{The ACA's Individual Mandate} --- Under the ACA, most Americans were required to maintain health coverage or face a financial penalty. If this mandate had more ``bite" in non-expansion states than expansion states, it might account for the divergence in original Medicaid enrollment between the two groups of states. However, this explanation is tenuous for two reasons: First, the mandate would essentially have augmented the ``woodwork effect" --- drawing even more eligible-but-not-enrolled people onto Medicaid. But as we discuss in Section \ref{woodwork_effect}, survey-based studies do not support the view that the ``woodwork" was substantially larger in non-expansion states. Second, since the individual mandate was enforced by the IRS, a federal agency, there is no reason to believe that residents of non-expansion states experienced more vigorous enforcement. Third, if the individual mandate had played an important role in causing the divergence in the growth of the original Medicaid population from 2014 to 2019, one would expect Figure \ref{figure1} to show accelerating growth in both expansion and non-expansion states, with non-expansion states rising faster. In reality, the divergence stems from stagnating growth among expansion states, not particularly rapid growth in non-expansion states.

\vspace{0.5cm}
\noindent \textit{``Silver Loading"} --- In the fall of 2017, the Trump administration stopped reimbursing health insurers operating in the ACA's exchanges for cost-sharing reductions (CSRs) that certain lower-income households are entitled to. The decision caused temporary disruption to the individual health insurance market, but ultimately resulted in lower premiums for millions of consumers on the exchanges as insurers built the cost of CSRs into premiums, triggering larger PTCs \citep{aron2019data,fiedler2021}. By making exchange coverage more affordable, ``silver loading" may have led some people to transition off Medicaid and on to private plans. We consider this implausible. First, the cessation of federal CSR payments that became the impetus for ``silver loading" did not occur until October 2017. Therefore, it cannot explain the clear effects we find in 2016 and 2017 (see Figure \ref{appendix_2014_states_results_plot}).\footnote{Recall that our data on Medicaid enrollment represents the month of June in each year (see Section \ref{Data}), so our estimates for 2017 precede the elimination of federal CSR payments.} Second, as mentioned above, individuals eligible for Medicaid are not eligible for PTCs, so this hypothesis assumes that millions of households reacted to ``silver loading" by altering their income or other characteristics to become eligible for PTCs. Moreover, \citet{aron2019data} notes that ``silver loading" was least beneficial for people with incomes between 100 percent and 200 percent of the poverty level --- implying that the group for whom such strategic behavior may have  been the most feasible had the least incentive to do so. Third, this hypothesis requires ``silver loading" to have had a substantially larger effect on Medicaid enrollment in expansion states than in non-expansion states. Based on state-level estimates of the number of consumers affected by ``silver-loading" \citep{aron2018}, we see little evidence that this was the case.

\subsection{Fiscal Impact of Reclassifications}
Using our estimates of the effects of Medicaid expansion on enrollment in the original Medicaid population, we now turn to back-of-the-envelope calculations of its fiscal impact on states and the federal government. For the purposes of deriving quantitative fiscal estimates, we assume that all those who would otherwise have been enrolled in the original Medicaid population were reclassified into the new adult group.\footnote{Despite this assumption, we likely still underestimate the number of original Medicaid enrollees reclassified to the new adult group, since research using household survey data indicates that the woodwork effect induced by the ACA was larger in expansion states than in non-expansion states. As a result, non-expansion states are likely to underestimate the counterfactual level of enrollment in the original Medicaid population in expansion states. See Section \ref{woodwork_effect} for more details.} Because of the difference in FMAP rates applicable to the original Medicaid population and the new adult group, reclassifications represent a substantial federal subsidy to states. We approximate the size of the subsidy for each state and year using
\begin{equation}
    \textit{Y}_{i,t} = \text{estimated enrollees reclassified}_{i,t} \times \text{FMAP rate spread}_{i,t} \times \text{per enrollee expenditures}_t
\end{equation}
\noindent where \(Y_{i,t}\) represents the reclassification-related Medicaid subsidy received by state \(i\) in year \(t\), \textit{estimated enrollees reclassified} represents the difference between actual enrollment in the original Medicaid population and our estimated counterfactual enrollment,\footnote{We use our coefficient from column (3) in Table \ref{tab:main_results} (--0.0993) to derive the counterfactual enrollment levels in each state. To do so, we multiply actual enrollment in a given state and year by \(\frac{1}{1-0.0993}\) = 1.1102.} \textit{FMAP rate spread} is the difference between the traditional FMAP rate and the enhanced FMAP rate,\footnote{Over our sample period, the mean FMAP rate spread among expansion states was 0.39; the median was 0.43.} and \textit{per enrollee expendiutures} is the national average of expenditures per non-elderly adult Medicaid enrollee (expressed in constant 2019 dollars), excluding the new adult group.\footnote{We obtain per enrollee expenditures from annual reports compiled by the Medicaid and CHIP Payment and Access Commission (MACPAC). Estimates are available for 2013, 2018, and 2019. Estimates were not published for 2014, 2015, 2016 or 2017. To estimate per enrollee expenditures in the missing years, we perform a linear interpolation using 2013 and 2018 as endpoints. We aggregate per enrollee expenditures up to the national level because of data quality concerns with state-level estimates. All years are converted to 2019 dollars using the CPI. The estimated annual per enrollee expenditures (in 2019 dollars) rose from \$4,612 in 2014 to \$4,908 in 2019.}

Our results are presented in Table \ref{subsidy_estimates_table}. The fiscal impact of the reclassifications we document is substantial. Our estimates imply that \$52.9 billion in additional federal funding was distributed to states from 2014 to 2019 based on these reclassifications. Over that period, approximately 26.2 million reclassifications (measured as enrollee-years) may have occurred across all expansion states. Since Medicaid expansion was adopted in a staggered fashion over our sample period, however, the cumulative totals are somewhat distorted by the fact that some states expanded Medicaid in later years. To address this, Table \ref{subsidy_estimates_table} also shows estimates for 2019, the last year in our sample.\footnote{Despite more states belonging to the expansion cohort in 2019 than in previous years, the total state subsidy in 2019 (\$8.3 billion) is slightly smaller than the average annual subsidy (\(\frac{\mbox{\$52.9}}{\mbox{6 years}}\) = \$8.8 billion) over the 2014-2019 period because the enhanced FMAP rate declined from 100\% in 2014 to 93\% in 2019.} That year, the original Medicaid population had approximately 4.4 million fewer beneficiaries as a result of expansion, resulting in \$8.3 billion in subsidies to states, assuming these were reclassified into the new adult group. For context, federal Medicaid expenditures totaled \$405 billion in fiscal year 2019. Therefore, we calculate that the ACA's hidden subsidy may have accounted for approximately 2.0\% of federal Medicaid outlays that year.

\begin{table}[H]
\caption{\label{subsidy_estimates_table}Estimated State Subsidies from Reclassifications}
\centering
\begin{threeparttable}
\scriptsize
\begin{tabular}{lrrrr}
\hline \hline
 & \textbf{(1)} & \textbf{(2)} & \textbf{(3)} & \textbf{(4)} \\
\textbf{State} & \textbf{Enrollees Reclassified} & \textbf{Subsidy (Millions of \$,} & \textbf{Enrollees Reclassified} & \textbf{Subsidy (Millions of \$,} \\
\\[-6ex]
      & \textbf{(2014-2019)} &  \textbf{2014-2019)} & \textbf{(2019)} & \textbf{2019)} \\
\\[-4ex]
\hline
Alaska               & 65,197                         & \$            143,908,236  & 16,898                        & \$            35,661,429 \\
Arizona              & 1,143,180                      & \$         1,532,214,257   & 193,490                       & \$          220,223,764  \\
Arkansas             & 415,174                        & \$            531,401,967  & 63,461                        & \$            70,049,185 \\
California           & 6,301,045                      & \$       14,249,028,204    & 964,592                       & \$       2,035,712,772   \\
Colorado             & 578,605                        & \$         1,295,730,016   & 94,860                        & \$          200,195,982  \\
Connecticut          & 451,198                        & \$         1,015,806,169   & 78,068                        & \$          164,758,053  \\
Delaware             & 125,772                        & \$            251,946,480  & 18,527                        & \$            32,234,178 \\
District of Columbia & 123,553                        & \$            159,419,572  & 21,262                        & \$            24,001,881 \\
Hawaii               & 189,060                        & \$            391,952,540  & 31,090                        & \$            59,632,187 \\
Illinois             & 1,490,936                      & \$         3,319,867,980   & 222,405                       & \$          465,988,054  \\
Indiana              & 556,220                        & \$            810,178,691  & 109,948                       & \$          145,914,499  \\
Iowa                 & 296,977                        & \$            565,883,377  & 49,309                        & \$            80,032,042 \\
Kentucky             & 558,725                        & \$            706,783,889  & 92,468                        & \$            96,802,505 \\
Louisiana            & 381,252                        & \$            565,793,857  & 123,108                       & \$          169,180,471  \\
Maine                & 27,384                         & \$              38,277,057 & 27,384                        & \$            38,277,057 \\
Maryland             & 600,868                        & \$         1,353,645,229   & 101,313                       & \$          213,815,775  \\
Massachusetts        & 1,148,144                      & \$         2,592,755,094   & 160,450                       & \$          338,620,310  \\
Michigan             & 1,138,343                      & \$         1,731,004,370   & 193,986                       & \$          271,820,305  \\
Minnesota            & 600,981                        & \$         1,356,606,456   & 94,384                        & \$          199,192,440  \\
Montana              & 69,418                         & \$            101,424,056  & 17,304                        & \$            23,321,446 \\
Nevada               & 252,055                        & \$            390,746,502  & 41,558                        & \$            57,375,621 \\
New Hampshire        & 74,226                         & \$            166,749,362  & 14,187                        & \$            29,940,195 \\
New Jersey           & 735,045                        & \$         1,658,273,490   & 115,925                       & \$          244,653,227  \\
New Mexico           & 389,011                        & \$            487,757,313  & 62,712                        & \$            63,835,701 \\
New York             & 3,871,472                      & \$         8,735,439,467   & 635,420                       & \$       1,341,015,875   \\
North Dakota         & 47,045                         & \$            106,003,360  & 7,721                         & \$            16,294,819 \\
Ohio                 & 1,579,154                      & \$         2,601,153,570   & 246,572                       & \$          361,963,879  \\
Oregon               & 397,494                        & \$            635,702,735  & 63,689                        & \$            95,151,371 \\
Pennsylvania         & 1,146,557                      & \$         2,466,059,165   & 233,785                       & \$          467,571,680  \\
Rhode Island         & 150,946                        & \$            332,561,427  & 25,818                        & \$            51,230,964 \\
Vermont              & 122,864                        & \$            253,484,030  & 18,352                        & \$            35,226,675 \\
Virginia             & 137,448                        & \$            290,074,709  & 137,448                       & \$          290,074,709  \\
Washington           & 802,355                        & \$         1,808,644,578   & 131,438                       & \$          277,391,244  \\
West Virginia        & 246,826                        & \$            293,702,925  & 39,591                        & \$            36,258,941    \\
\hline
\textit{Total} & 26,214,527 & \$  52,939,980,131  & 4,448,523 & \$  8,253,419,233 \\ \hline\hline
& & & & \\[-6.25ex]
\end{tabular}
\begin{tablenotes}
    \item \scriptsize \textit{Notes}: This table reports the estimated federal payments distributed to states based on the reclassification of Medicaid enrollees from the original population to the new adult group. Authors' calculations. See main text for details. We omit states that had not expanded by 2019, when our sample period ends.
\end{tablenotes}
\end{threeparttable}
\end{table}
\clearpage

Several strands of circumstantial evidence support the reclassification hypothesis. Despite fears that a large woodwork effect would put substantial strain on state budgets, subsequent analyses have revealed that the fiscal impact of Medicaid expansion has been smaller than anticipated \citep{sommers2017federal, gruber2020fiscal}, with some analyses appearing to show that Medicaid expansion resulted in net fiscal savings in some states \citep{levy2020macroeconomic,simpson2020implications}. Reclassifications, by allowing states to blunt the woodwork effect and draw down additional federal Medicaid funding through the enhanced FMAP rate, help to explain this outcome. Relatedly, enrollment in the new adult group enrollment has exceeded projections in virtually every expansion state \citep{blase2019}. Reclassifications, which were generally not contemplated by forecasters, provide a simple explanation. Finally, per enrollee spending on the new adult group has been significantly higher than predicted. In 2013, CMS estimated that per enrollee costs in the new adult group would be \$3,625 per enrollee in 2016 \citep{cms2013}. A subsequent report from the same source revealed that members of the new adult group had, in fact, cost \$5,959 per enrollee in 2016 \citep{cms2018}, nearly two-thirds more than originally predicted. This fact is consistent with the notion that some of the original Medicaid population --- who are more costly to insure, on average, than members of the new adult group --- were reclassified into the new adult group.

Our data provides little direct insight into the types of enrollees being reclassified. As we discussed in Section \ref{Reclassifications}, the ACA and subsequent federal rulemaking established some pathways whereby certain individuals who would otherwise have been enrolled in the original Medicaid population could be counted in the new adult group. For example, a woman who enrolls in Medicaid under the ACA rules and later becomes pregnant (thereby meeting eligibility criteria for the original Medicaid population) need not be reclassified into the original Medicaid population during her pregnancy. Similar logic applies to a person who enrolls in the new adult group and subsequently suffers a disabling injury that renders them eligible for Medicaid coverage under pre-ACA eligibility rules; they need not be transferred to the original Medicaid population for the purposes of obtaining federal reimbursements. These forces would tend to reduce the size of the original Medicaid population gradually, as more and more members of the new adult group experienced these health events. Yet our results are inconsistent with this prediction. The dynamic treatment effects we estimate indicate that the original Medicaid population contracted suddenly in the second post-treatment year and remained relatively stable over the succeeding four years, rather than continuing to decline. Therefore, we conclude that the forces ``siphoning off" enrollees from the original Medicaid population likely play a minor role in explaining our results.

On the other hand, our findings may partly be driven by reclassifications that occurred in violation of Medicaid rules. Under federal law, states are responsible for determining applicants’ eligibility for Medicaid, including periodically redetermining eligibility, disenrolling individuals who are no longer eligible, and reclassifying enrollees who may no longer meet the criteria under one eligibility pathway but may still qualify for Medicaid coverage through a different pathway. Yet the enhanced FMAP rates for the new adult group offered under the ACA dramatically reduced states’ incentives to maintain accurate Medicaid rolls. Moreover, the federal government provides only token oversight of states' eligibility verification procedures. According to CMS, ``When states submit their Medicaid expenditure reports, they certify the data are accurate and CMS conducts a limited review to assess whether the data is reasonable. The review consists of comparing the state-reported data to other readily available information, including state-reported performance indicators and expenditures, and follow-up with the state as needed.” Yet states rarely face meaningful penalties for submitting incorrect enrollment records. During the entire post-treatment period we examine (2014-2019), it was the explicit policy of the federal government not to attempt to recoup funds distributed to states on the basis of eligibility errors \citep{gao2020}. Previous research has noted that the ACA's Medicaid expansion was associated with large increases in Medicaid coverage among adults with incomes above 138\% of the federal poverty level, suggesting that states failed to adequately enforce eligibility rules \citep{courtemanche2019medicaid}.

Recent federal investigations into expansion states' Medicaid records provide direct evidence that improper reclassifications into the new adult group are common. In an audit of New York’s Medicaid program, investigators reviewed eligibility documentation for a random sample of 130 Medicaid enrollees New York had classified as belonging to the new adult group, and for whom New York had received funding through the ACA's enhanced FMAP rate. The audit found that New York incorrectly claimed enhanced reimbursement for 13.8\% of these enrollees and did not provide sufficient documentation to verify that 1.5\% of these enrollees were eligible for enhanced Medicaid reimbursement \citep{oig2019ny}. A similar audit in California found that 18.0\% of a randomly selected sample of enrollees in the new adult group were ineligible and 9.3\% of enrollees were potentially ineligible under ACA rules \citep{oig2018}. In Colorado, an investigation found that 23.3\% of randomly selected enrollees in the new adult group were ineligible, while an additional 6.7\% lacked sufficient documentation to determine eligibility \citep{oig2019}. 

Following \citet{bundorf2022responsiveness}, we extrapolate from these audits to provide a general indication of the proportion of reclassifications that may be improper. To do so, we use the results of the New York audit as a lower bound (using only the proportion of enrollees -- 13.8\% -- auditors verified as ineligible) and the results of the Colorado audit as an upper bound (using the proportion of enrollees -- 28.3\% -- auditors found to be definitely or potentially ineligible).\footnote{Out of the 60 Medicaid beneficiaries sampled, 14 were ineligible and 4 may have been ineligible, but one person was counted in both groups.} In 2019, states reported total enrollment in the new adult group of 12.0 million. Applying these lower and upper bounds, we find that between 1.65 million and up to 3.4 million of these enrollees may have been improper. In light of our finding that the original Medicaid population declined by 4.4 million enrollees, these figures suggest that between 37.2\% and 76.3\% of all reclassified enrollees may have been reported in violation of federal law. These estimates should be interpreted cautiously, however, since enrollment patterns may vary by state and over time; other expansion states may have higher or lower misclassification rates than New York, Colorado, or California.

\subsection{Robustness Tests} \label{robustness}

In this section, we perform several of tests to assess the sensitivity of our findings.

\subsubsection{Data Quality}

As explained in greater detail in Section \ref{Data}, our main analysis uses two different data sources to measure state Medicaid enrollment; one covers the years 2006 to 2013, while the other covers the years 2014 to 2019. Since most expansion states began implementing the reform in 2014, it is conceivable that our findings could be an artifact of transitioning to a different data source. This could occur if our 2014-2019 data systematically under-counted original Medicaid enrollment in expansion states relative to non-expansion states. The lack of a clear discontinuity between expansion and non-expansion states in 2014 --- visible in Figures \ref{figure1}, \ref{main_results_plot}, and \ref{appendix_2014_states_results_plot} --- is re-assuring. Still, we further explore this possibility in two ways. First, we re-run the analysis using only 2014-19 data. While this restricts our sample and limits the number of pre-treatment periods available, it obviates the need to combine different data sources. Results of that exercise are presented in Table \ref{tab:2014_weighted}. The treatment effect in our preferred specification (column 3) remains statistically significant, albeit somewhat smaller in magnitude (-0.0764 instead of -0.0993). Most specifications are no longer statistically significant at the 10\% level, and estimated effect sizes generally shrink compared to our main results. The small number of observations in our restricted sample may contribute to a loss of statistical significance. Still, we note that all coefficients remain negative and economically meaningful.

\begin{table}[H]
\centering
\caption{Effects of Medicaid Expansion on Enrollment: 2014-2019 Sample Period}
\label{tab:2014_weighted}
\begin{threeparttable}
\scriptsize
\begin{tabular}{l c c c c c c c c}
\toprule
 & (1) & (2) & (3) & (4) & (5) & (6) & (7) & (8) \\
\midrule
ATT & -0.0813\textsuperscript{**} & -0.0764\textsuperscript{**} & -0.0604\textsuperscript{} & -0.0526\textsuperscript{} & -0.0744\textsuperscript{**} & -0.0480\textsuperscript{} & -0.0651\textsuperscript{} & -0.0367\textsuperscript{} \\
 & (0.0334) & (0.0339) & (0.0417) & (0.0354) & (0.0330) & (0.0419) & (0.0408) & (0.0410)\\
\addlinespace \midrule
Gov's Political Party & & \checkmark & \checkmark & \checkmark & \checkmark & \checkmark & \checkmark & \checkmark \\
Elig. Limit, Parents & & \checkmark & & \checkmark & \checkmark & \checkmark & \checkmark & \checkmark \\
Unemployment Rate & & \checkmark & \checkmark & & \checkmark & \checkmark & \checkmark & \\
Elig. Limit, Children & & & \checkmark & & & & & \\
Poverty Rate & & & & \checkmark & & & & \\
ln(State Population) & & & & & \checkmark & & & \\
Non-White (\% of State) & & & & & & \checkmark & & \\
TANF Benefits & & & & & & & \checkmark & \\
Food Insecurity Rate & & & & & & & & \checkmark \\
\addlinespace \midrule
N & 150 & 150 & 150 & 150 & 150 & 150 & 150 & 150 \\
\bottomrule
& & & & & & & \\[-6.25ex]
\end{tabular}
\begin{tablenotes}
\footnotesize 
    \item \textit{Notes}: This table shows estimates of the ATT of Medicaid expansion on the size of the original Medicaid population (logged) across a range of models, all of which use the difference-in-differences estimator described by \citet{callaway2021difference} weighted by each state's 2013 Medicaid population. The comparison group is "never" treated units. All other models include all states. Standard errors (clustered by state) are reported in parentheses. \\* $p<0.10$, ** $p<0.05$, *** $p<0.01$.
\end{tablenotes}
\end{threeparttable}

\end{table}

As an additional check, we compare our 2013 Medicaid enrollment data from KFF with estimates from CMS of 2013 Medicaid and CHIP enrollment.\footnote{$~$The CMS estimates reflect the average monthly Medicaid and CHIP enrollment from July to September 2013. Our KFF estimates reflect Medicaid enrollment in June 2013. CMS did not release 2013 estimates for Medicaid only.} The CMS data, which has been used as a benchmark to gauge ACA-induced changes in coverage, is not available for prior years, but this narrow overlap in 2013 provides some insight into whether KFF and CMS estimates systematically differ. Figure \ref{data_check_scatterplot} in the Appendix plots the log of enrollment in each state from KFF and CMS in 2013. Nearly all states lie very close to the diagonal, indicating no large differences between the two sources. To the extent that some states lie slightly above the diagonal, this may be due to the fact that CMS' data includes CHIP enrollees, while KFF's estimates exclude these enrollees. We also note the absence of any clear pattern between expansion states (in blue) and non-expansion states (in red). Overall, these results assuage concerns that the our main findings are driven by data discrepancies.

\subsubsection{Changes to Enrollment Practices}

During our study period, some states implemented reforms to their administrative procedures that may have reduced enrollment in the original Medicaid population. In particular, \citet{arbogast2024administrative} document two major categories of new rules: 1) increases in the stringency and frequency of eligibility and income checks, and 2) mechanisms to automatically disenroll beneficiaries deemed to no longer qualify for the program (e.g., cancelling someone’s coverage without notice if a person does not respond to a request for documentation within a certain time frame). To the extent that these policies coincided with Medicaid expansion and may have disproportionately affected populations in expansion states, they could influence our findings. To address this concern, we re-estimate our main models after dropping the 13 states that implemented one or both of these policies from 2013-2019. The results of this exercise, which we report in Table \ref{tab:adminburdens}, are generally similar to our main estimates; all specifications that reached statistical significance in our main analysis remain statistically significant and some coefficients — including our preferred specification — grow slightly in magnitude. The dynamic treatment effects we obtain from this more limited subset of states, shown in Figure \ref{event_plot_adminburdens}, are also similar to our main results.

\vspace{0.5cm}

\begin{table}[H]
\centering
\caption{Effects of Medicaid Expansion on Enrollment: Exclude Observations from States That Imposed Administrative Burdens}
\label{tab:adminburdens}
\begin{threeparttable}
\scriptsize
\begin{tabular}{l c c c c c c c c}
\toprule
 & (1) & (2) & (3) & (4) & (5) & (6) & (7) & (8) \\
\midrule
ATT (S.E.) & -0.0299\textsuperscript{} & -0.1031\textsuperscript{**} & -0.0654\textsuperscript{**} & -0.1086\textsuperscript{**} & -0.0950\textsuperscript{*} & -0.0943\textsuperscript{*} & -0.0607\textsuperscript{*} & -0.0749\textsuperscript{**} \\
 & (0.0383) & (0.0404) & (0.0332) & (0.0429) & (0.0499) & (0.0488) & (0.0339) & (0.0378)\\
\addlinespace \midrule
Gov's Political Party & & \checkmark & \checkmark & \checkmark & \checkmark & \checkmark & \checkmark & \checkmark \\
Elig. Limit, Parents & & \checkmark & & \checkmark & \checkmark & \checkmark & \checkmark & \checkmark \\
Unemployment Rate & & \checkmark & \checkmark & & \checkmark & \checkmark & \checkmark & \\
Elig. Limit, Children & & & \checkmark & & & & & \\
Poverty Rate & & & & \checkmark & & & & \\
ln(State Population) & & & & & \checkmark & & & \\
Non-White (\% of State) & & & & & & \checkmark & & \\
TANF Benefits & & & & & & & \checkmark & \\
Food Insecurity Rate & & & & & & & & \checkmark \\
\addlinespace \midrule
N & 669 & 669 & 669 & 669 & 669 & 669 & 669 & 669 \\
\bottomrule
& & & & & & & & \\[-6.25ex]
\end{tabular}
\begin{tablenotes}
\footnotesize 
    \item \textit{Notes}: This table shows estimates of the ATT of Medicaid expansion on the size of the original Medicaid population (logged) across a range of models, all of which use the difference-in-differences estimator described by \citet{callaway2021difference} excluding observations from states that had imposed administrative burdens of more stringent eligibility checks (TX, MS, LA, ID, IL, HI, FL, CO) and states that implemented automatic disenrollment policies during our sample period (AR, IL, LA, MO, NC, OH, TN). For more, see \cite{arbogast2024administrative}. The comparison group is "never" treated states. Standard errors (clustered by state) are reported in parentheses. \\
* $p<0.10$, ** $p<0.05$, *** $p<0.01$.
\end{tablenotes}
\end{threeparttable}
\end{table}

\bigskip\bigskip

\setcounter{figure}{3}
\renewcommand{\thefigure}{\arabic{figure}}
\vspace{.5cm}
\begin{figure}[H]
\centering

\caption{\label{event_plot_adminburdens} Dynamic Effects of Medicaid Expansions on Enrollment: Exclude Observations from States That Imposed Administrative Burdens}
\includegraphics[width=\textwidth]{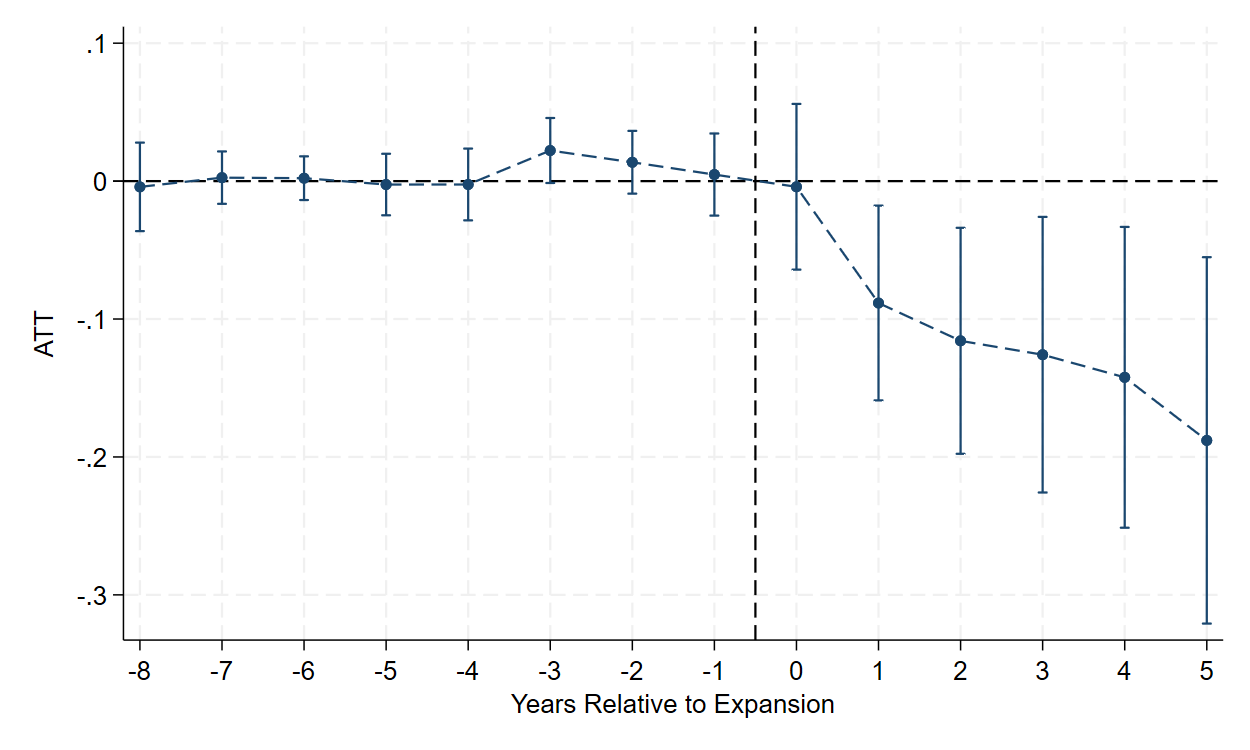}
\caption*{{\footnotesize \textit{Notes}: This plot shows dynamic effects across event time, from (column (2) in Table \ref{tab:adminburdens}), excluding observations from states that had imposed administrative burdens of more frequent/stringent eligibility checks (TX, MS, LA, ID, IL, HI, FL, CO) and states that implemented automatic disenrollment policies during our sample period (AR, IL, LA, MO, NC, OH, TN). For more, see \cite{arbogast2024administrative}. Bars represent 95\% confidence intervals. The vertical dashed line represents the implementation of Medicaid expansion. We use enrollment figures for the month of June in each year. Data was compiled by the authors from Kaiser Family Foundation issue briefs (for the years 2006-2013) and reports from the Medicaid Budget and Expenditure System (for the years 2014-2019); see section \ref{Data} for more details. We define the original Medicaid population as total Medicaid enrollment minus the number of enrollees reported by states as ``newly eligible" under the ACA.} }
\end{figure}

\section{Conclusion}

The expansion of Medicaid under the ACA was a significant development in U.S. health policy. We examine a previously overlooked fiscal effect of this reform. Although past research using household survey data has documented a robust woodwork effect in Medicaid associated with expansion, we find no evidence of such an effect in states' administrative enrollment records. Rather, we find evidence that the original Medicaid population contracted sharply following Medicaid expansion's implementation, defying forecasters' expectations. We argue that this discrepancy is a mirage caused by the reclassification of individuals who otherwise would have been counted in the original Medicaid population to the new adult group. While these reclassifications were a purely administrative phenomenon that did not affect the coverage or benefits of individual Medicaid enrollees, they have had substantial fiscal effects on states and the federal government.

Our estimates imply that these reclassifications resulted in nearly \$52.9 billion in federal Medicaid payments to states from 2014-2019, including \$8.3 billion in 2019 alone. The hidden subsidy we document represents a sizable share of Medicaid expansion's impact on federal spending. According to the CBO, the direct federal costs of Medicaid expansion --- that is, reimbursements made to states to cover medical services for the new adult group --- were \$66 billion in 2019 \citep{cbo2019}. This figure, however, implicitly assumes that members of the new adult group would not have received federal subsidies in the absence of the ACA's expanded eligibility rules. Our results indicate, however, that 4.4 million Medicaid enrollees classified in the new adult group may have been counted as original Medicaid enrollees and reimbursed at states' traditional FMAP rate if Medicaid expansion had not occurred. Our results imply that the federal government may have provided \$12.1 billion to states in 2019 to cover these enrollees.\footnote{We arrive at this result by multiplying the estimated number of reclassified Medicaid enrollees in 2019 in each state (totalling 4.4 million) by the average per enrollee cost nationwide in 2019 (\$4,908) and the applicable state's traditional FMAP rate.} Therefore, the federal fiscal impact of Medicaid expansion is approximately \$53.9 billion (\$66 billion -- \$12.1 billion), substantially smaller than CBO's estimates suggest. On the other hand, these downward revisions imply that reclassifications inflated the federal cost of Medicaid expansion by 18.2\%.

It is likely that similar subsidies occurred in more recent years, although forces linked to the COVID-19 public health emergency may have changed their magnitude. On the one hand, the temporary increase in the traditional FMAP rate during the COVID-19 public health emergency narrowed the FMAP rate spread with the enhanced FMAP rate, which would have reduced the size of the hidden subsidy. On the other hand, the continuous enrollment requirement imposed during the COVID-19 public health emergency substantially increased Medicaid enrollment and may have increased the size of the hidden subsidy.

Our results suggest that state policymakers are sensitive to incentives created through Medicaid's joint financing structure. The ACA's hidden subsidy has had a substantial fiscal effect on the federal government and expansion states. Accounting for strategic behavior by states is crucial for accurately predicting the effects of policy changes to Medicaid and similar federal-state programs. More stringent federal monitoring of states' enrollment practices may help to mitigate such behavior. Alternative financing methods, such as federal block grants, could also reduce or eliminate opportunities to draw down additional federal Medicaid funding through administrative reclassifications.

\clearpage
\singlespacing
\bibliographystyle{ecta}
\bibliography{references}

\begin{thebibliography}{60}
\newcommand{\enquote}[1]{``#1''}
\expandafter\ifx\csname natexlab\endcsname\relax\def\natexlab#1{#1}\fi

\bibitem[\protect\citeauthoryear{Adams and Wade}{Adams and
  Wade}{2001}]{adams2001fiscal}
\textsc{Adams, E.~K. and M.~Wade} (2001): \enquote{Fiscal response to a
  matching grant: Medicaid expenditures and enrollments, 1984-1992,}
  \emph{Public Finance Review}, 29, 26--48.

\bibitem[\protect\citeauthoryear{Arbogast, Chorniy, and Currie}{Arbogast
  et~al.}{2024}]{arbogast2024administrative}
\textsc{Arbogast, I., A.~Chorniy, and J.~Currie} (2024):
  \enquote{Administrative Burdens and Child Medicaid and CHIP Enrollments,}
  \emph{American Journal of Health Economics}, 10, 237--271.

\bibitem[\protect\citeauthoryear{Arenberg, Neller, and Stripling}{Arenberg
  et~al.}{2024}]{arenberg2024impact}
\textsc{Arenberg, S., S.~Neller, and S.~Stripling} (2024): \enquote{The impact
  of youth Medicaid eligibility on adult incarceration,} \emph{American
  Economic Journal: Applied Economics}, 16, 121--156.

\bibitem[\protect\citeauthoryear{Aron-Dine}{Aron-Dine}{2018}]{aron2018}
\textsc{Aron-Dine, A.} (2018): \enquote{Individual Market Stabilization
  Proposals Should Avoid Raising Costs for Consumers,} Tech. rep., Center for
  Budget and Policy Priorities.

\bibitem[\protect\citeauthoryear{Aron-Dine}{Aron-Dine}{2019}]{aron2019data}
---\hspace{-.1pt}---\hspace{-.1pt}--- (2019): \enquote{Data: silver loading is
  boosting insurance coverage,} \emph{Health Affairs Forefront}.

\bibitem[\protect\citeauthoryear{Artiga and Stephens}{Artiga and
  Stephens}{2013}]{artiga2013}
\textsc{Artiga, S. and J.~Stephens} (2013): \enquote{Key Lessons from Medicaid
  andCHIP for Outreach and EnrollmentUnder the Affordable Care Act,} Tech.
  rep., Kaiser Family Foundation.

\bibitem[\protect\citeauthoryear{Artiga, Tolbert, and Rudowitz}{Artiga
  et~al.}{2015}]{artiga2015}
\textsc{Artiga, S., J.~Tolbert, and R.~Rudowitz} (2015): \enquote{Year Two of
  the ACA Coverage Expansions: On-the-Ground Experiences from Five States,}
  Tech. rep., Kaiser Family Foundation.

\bibitem[\protect\citeauthoryear{Baicker}{Baicker}{2005}]{baicker2005spillover}
\textsc{Baicker, K.} (2005): \enquote{The spillover effects of state spending,}
  \emph{Journal of Public Economics}, 89, 529--544.

\bibitem[\protect\citeauthoryear{Bartik}{Bartik}{2002}]{bartik2002spillover}
\textsc{Bartik, T.~J.} (2002): \enquote{Spillover effects of welfare reforms in
  state labor markets,} \emph{Journal of Regional Science}, 42, 667--701.

\bibitem[\protect\citeauthoryear{Blase}{Blase}{2016}]{blase2016}
\textsc{Blase, B.} (2016): \enquote{Evidence Is Mounting: The Affordable Care
  Act Has Worsened Medicaid’s Structural Problems,} Tech. rep., Mercatus
  Center at George Mason University.

\bibitem[\protect\citeauthoryear{Blase and Albanese}{Blase and
  Albanese}{2022}]{blase2022}
\textsc{Blase, B. and J.~Albanese} (2022): \enquote{America’s Largest Health
  Care Programs Are Full of Improper Payments,} Tech. rep., Paragon Health
  Institute.

\bibitem[\protect\citeauthoryear{Blase and Yelowitz}{Blase and
  Yelowitz}{2019}]{blase2019}
\textsc{Blase, B. and A.~Yelowitz} (2019): \enquote{The ACA’s Medicaid
  Expansion: A Review of Ineligible Enrollees and Improper Payments,} Tech.
  rep., Mercatus Center at George Mason University.

\bibitem[\protect\citeauthoryear{Borusyak, Jaravel, and Spiess}{Borusyak
  et~al.}{2021}]{borusyak2021revisiting}
\textsc{Borusyak, K., X.~Jaravel, and J.~Spiess} (2021): \enquote{Revisiting
  event study designs: Robust and efficient estimation,} \emph{arXiv preprint
  arXiv:2108.12419}.

\bibitem[\protect\citeauthoryear{Buchmueller, Miller, and Vujicic}{Buchmueller
  et~al.}{2016}]{buchmueller2016providers}
\textsc{Buchmueller, T., S.~Miller, and M.~Vujicic} (2016): \enquote{How do
  providers respond to changes in public health insurance coverage? Evidence
  from adult Medicaid dental benefits,} \emph{American Economic Journal:
  Economic Policy}, 8, 70--102.

\bibitem[\protect\citeauthoryear{Bundorf and Kessler}{Bundorf and
  Kessler}{2022}]{bundorf2022responsiveness}
\textsc{Bundorf, M.~K. and D.~P. Kessler} (2022): \enquote{The responsiveness
  of medicaid spending to the federal subsidy,} \emph{National Tax Journal},
  75, 661--680.

\bibitem[\protect\citeauthoryear{Butler}{Butler}{2016}]{butler2016future}
\textsc{Butler, S.~M.} (2016): \enquote{The future of the Affordable Care Act:
  Reassessment and revision,} \emph{JAMA}, 316, 495--497.

\bibitem[\protect\citeauthoryear{Callaway and Sant’Anna}{Callaway and
  Sant’Anna}{2021}]{callaway2021difference}
\textsc{Callaway, B. and P.~H. Sant’Anna} (2021):
  \enquote{Difference-in-differences with multiple time periods,} \emph{Journal
  of Econometrics}, 225, 200--230.

\bibitem[\protect\citeauthoryear{Carey, Miller, and Wherry}{Carey
  et~al.}{2020}]{carey2020impact}
\textsc{Carey, C.~M., S.~Miller, and L.~R. Wherry} (2020): \enquote{The impact
  of insurance expansions on the already insured: the Affordable Care Act and
  Medicare,} \emph{American Economic Journal: Applied Economics}, 12, 288--318.

\bibitem[\protect\citeauthoryear{{Centers for Medicare \& Medicaid
  Services}}{{Centers for Medicare \& Medicaid Services}}{2019}]{cms2019perm}
\textsc{{Centers for Medicare \& Medicaid Services}} (2019): \enquote{Payment
  Error Rate Measurement (PERM) Program Medicaid Improper Payment Rates,} Tech.
  rep., Centers for Medicaid \& Medicaid Services.

\bibitem[\protect\citeauthoryear{Chiedi}{Chiedi}{2018}]{oig2019ny}
\textsc{Chiedi, J.} (2018): \enquote{New York Incorrectly Claimed Enhanced
  Federal Medicaid Reimbursement for Some Beneficiaries,} Tech. rep.,
  Department of Health and Human Services, Office of Inspector General,
  A-02-15-01023.

\bibitem[\protect\citeauthoryear{Chiedi}{Chiedi}{2019}]{oig2019}
---\hspace{-.1pt}---\hspace{-.1pt}--- (2019): \enquote{Colorado Did Not
  Correctly Determine Medicaid Eligibility For Some Newly Enrolled
  Beneficiaries,} Tech. rep., Department of Health and Human Services, Office
  of Inspector General, A-07-16-04228.

\bibitem[\protect\citeauthoryear{Courtemanche, Marton, and
  Yelowitz}{Courtemanche et~al.}{2019}]{courtemanche2019medicaid}
\textsc{Courtemanche, C.~J., J.~Marton, and A.~Yelowitz} (2019):
  \enquote{Medicaid coverage across the income distribution under the
  Affordable Care Act,} Tech. rep., National Bureau of Economic Research.

\bibitem[\protect\citeauthoryear{De~Chaisemartin and
  d’Haultfoeuille}{De~Chaisemartin and d’Haultfoeuille}{2020}]{de2020two}
\textsc{De~Chaisemartin, C. and X.~d’Haultfoeuille} (2020): \enquote{Two-way
  fixed effects estimators with heterogeneous treatment effects,}
  \emph{American Economic Review}, 110, 2964--2996.

\bibitem[\protect\citeauthoryear{De~La~Mata}{De~La~Mata}{2012}]{de2012effect}
\textsc{De~La~Mata, D.} (2012): \enquote{The effect of Medicaid eligibility on
  coverage, utilization, and children's health,} \emph{Health Economics}, 21,
  1061--1079.

\bibitem[\protect\citeauthoryear{Fiedler}{Fiedler}{2021}]{fiedler2021}
\textsc{Fiedler, M.} (2021): \enquote{The case for replacing `Silver Loading',}
  Tech. rep., Brookings Institution.

\bibitem[\protect\citeauthoryear{Frank}{Frank}{2014}]{aspe_brief2014}
\textsc{Frank, R.} (2014): \enquote{Health Insurance Marketplace: Summary of
  Enrollment Report for the Initial Annual Open Enrollment Period,} Tech. rep.,
  U.S. Department of Health and Human Services, Office of the Assistant
  Secretary for Planning and Evaluation.

\bibitem[\protect\citeauthoryear{Frean, Gruber, and Sommers}{Frean
  et~al.}{2017}]{frean2017premium}
\textsc{Frean, M., J.~Gruber, and B.~D. Sommers} (2017): \enquote{Premium
  subsidies, the mandate, and Medicaid expansion: Coverage effects of the
  Affordable Care Act,} \emph{Journal of Health Economics}, 53, 72--86.

\bibitem[\protect\citeauthoryear{Fritzsche, McNellis, and Vreeland}{Fritzsche
  et~al.}{2019}]{cbo2019}
\textsc{Fritzsche, K., K.~McNellis, and E.~Vreeland} (2019): \enquote{Federal
  Subsidies for Health Insurance Coverage for People Under Age 65: 2019 to
  2029,} Tech. rep., Congressional Budget Office.

\bibitem[\protect\citeauthoryear{Goodman-Bacon}{Goodman-Bacon}{2021}]{goodman2021difference}
\textsc{Goodman-Bacon, A.} (2021): \enquote{Difference-in-differences with
  variation in treatment timing,} \emph{Journal of Econometrics}, 225,
  254--277.

\bibitem[\protect\citeauthoryear{Grabowski}{Grabowski}{2006}]{grabowski2006cost}
\textsc{Grabowski, D.~C.} (2006): \enquote{The cost-effectiveness of
  noninstitutional long-term care services: Review and synthesis of the most
  recent evidence,} \emph{Medical Care Research and Review}, 63, 3--28.

\bibitem[\protect\citeauthoryear{Grannemann and Pauly}{Grannemann and
  Pauly}{1983}]{grannemann1983controlling}
\textsc{Grannemann, T.~W. and M.~V. Pauly} (1983): \emph{Controlling Medicaid
  Costs: Federalism, Competition, and Choice}, Washington: AEI: American
  Enterprise Institute for Public Policy Research.

\bibitem[\protect\citeauthoryear{Gruber and Sommers}{Gruber and
  Sommers}{2019}]{gruber2019affordable}
\textsc{Gruber, J. and B.~D. Sommers} (2019): \enquote{The Affordable Care
  Act's effects on patients, providers, and the economy: what we've learned so
  far,} \emph{Journal of Policy Analysis and Management}, 38, 1028--1052.

\bibitem[\protect\citeauthoryear{Gruber and Sommers}{Gruber and
  Sommers}{2020}]{gruber2020fiscal}
---\hspace{-.1pt}---\hspace{-.1pt}--- (2020): \enquote{Fiscal federalism and
  the budget impacts of the Affordable Care Act's Medicaid expansion,} Tech.
  rep., National Bureau of Economic Research.

\bibitem[\protect\citeauthoryear{Hill, Wilkinson, and Courtot}{Hill
  et~al.}{2014}]{ian2014}
\textsc{Hill, I., M.~Wilkinson, and B.~Courtot} (2014): \enquote{The Launch of
  the Affordable Care Act in Selected States: Outreach, Education, and
  Enrollment Assistance,} Tech. rep., The Urban Institute.

\bibitem[\protect\citeauthoryear{Hudson and Moriya}{Hudson and
  Moriya}{2017}]{hudson2017medicaid}
\textsc{Hudson, J.~L. and A.~S. Moriya} (2017): \enquote{Medicaid expansion for
  adults had measurable ‘welcome mat’ effects on their children,}
  \emph{Health Affairs}, 36, 1643--1651.

\bibitem[\protect\citeauthoryear{Huh}{Huh}{2021}]{huh2021medicaid}
\textsc{Huh, J.} (2021): \enquote{Medicaid and provider supply,} \emph{Journal
  of Public Economics}, 200, 104430.

\bibitem[\protect\citeauthoryear{Imai and Kim}{Imai and
  Kim}{2021}]{imai2021use}
\textsc{Imai, K. and I.~S. Kim} (2021): \enquote{On the use of two-way fixed
  effects regression models for causal inference with panel data,}
  \emph{Political Analysis}, 29, 405--415.

\bibitem[\protect\citeauthoryear{Leung}{Leung}{2022}]{leung2022state}
\textsc{Leung, P.} (2022): \enquote{State responses to federal matching grants:
  The case of medicaid,} \emph{Journal of Public Economics}, 216, 104746.

\bibitem[\protect\citeauthoryear{Levinson}{Levinson}{2018}]{oig2018}
\textsc{Levinson, D.} (2018): \enquote{California Made Medicaid Payments on
  Behalf of Newly Eligible Beneficiaries Who Did Not Meet Federal and State
  Requirements,} Tech. rep., Department of Health and Human Services, Office of
  Inspector General, A-09-16-02023.

\bibitem[\protect\citeauthoryear{Levy, Ayanian, Buchmueller, Grimes, and
  Ehrlich}{Levy et~al.}{2020}]{levy2020macroeconomic}
\textsc{Levy, H., J.~Z. Ayanian, T.~C. Buchmueller, D.~R. Grimes, and
  G.~Ehrlich} (2020): \enquote{Macroeconomic feedback effects of Medicaid
  expansion: Evidence from Michigan,} \emph{Journal of Health Politics, Policy
  and Law}, 45, 5--48.

\bibitem[\protect\citeauthoryear{MACPAC}{MACPAC}{2017}]{macpac_state_medicaid_funds}
\textsc{MACPAC} (2017): \enquote{Medicaid’s share of state budgets,} Accessed
  on September 4, 2023.

\bibitem[\protect\citeauthoryear{McInerney, Mellor, and Sabik}{McInerney
  et~al.}{2017}]{mcinerney2017effects}
\textsc{McInerney, M., J.~M. Mellor, and L.~M. Sabik} (2017): \enquote{The
  effects of state Medicaid expansions for working-age adults on senior
  Medicare beneficiaries,} \emph{American Economic Journal: Economic Policy},
  9, 408--438.

\bibitem[\protect\citeauthoryear{McInerney, Mellor, and Sabik}{McInerney
  et~al.}{2021}]{mcinerney2021welcome}
---\hspace{-.1pt}---\hspace{-.1pt}--- (2021): \enquote{Welcome mats and
  on-ramps for older adults: The impact of the Affordable Care Act's Medicaid
  expansions on dual enrollment in Medicare and Medicaid,} \emph{Journal of
  Policy Analysis and Management}, 40, 12--41.

\bibitem[\protect\citeauthoryear{Miller, Johnson, and Wherry}{Miller
  et~al.}{2021}]{miller2021medicaid}
\textsc{Miller, S., N.~Johnson, and L.~R. Wherry} (2021): \enquote{Medicaid and
  mortality: new evidence from linked survey and administrative data,}
  \emph{The Quarterly Journal of Economics}, 136, 1783--1829.

\bibitem[\protect\citeauthoryear{Mitchell, Baumrucker, Colello, Napili, Binder,
  and Braun}{Mitchell et~al.}{2023}]{mitchell2023medicaid}
\textsc{Mitchell, A., E.~P. Baumrucker, K.~J. Colello, A.~Napili, C.~Binder,
  and S.~K. Braun} (2023): \enquote{Medicaid: An Overview,} Tech. rep.,
  Congressional Research Service.

\bibitem[\protect\citeauthoryear{Neprash, Zink, Sheridan, and
  Hempstead}{Neprash et~al.}{2021}]{neprash2021effect}
\textsc{Neprash, H.~T., A.~Zink, B.~Sheridan, and K.~Hempstead} (2021):
  \enquote{The effect of Medicaid expansion on Medicaid participation, payer
  mix, and labor supply in primary care,} \emph{Journal of Health Economics},
  80, 102541.

\bibitem[\protect\citeauthoryear{Nikpay}{Nikpay}{2022}]{nikpay2022medicaid}
\textsc{Nikpay, S.} (2022): \enquote{The Medicaid windfall: Medicaid expansions
  and the target efficiency of hospital safety-net subsidies,} \emph{Journal of
  Public Economics}, 208, 104583.

\bibitem[\protect\citeauthoryear{Peng}{Peng}{2017}]{peng2017does}
\textsc{Peng, L.} (2017): \enquote{How does medicaid expansion affect premiums
  in the health insurance marketplaces? New evidence from late adoption in
  Pennsylvania and Indiana,} \emph{American Journal of Health Economics}, 3,
  550--576.

\bibitem[\protect\citeauthoryear{Price and Saltzman}{Price and
  Saltzman}{2013}]{price2013economic}
\textsc{Price, C.~C. and E.~Saltzman} (2013): \enquote{The economic impact of
  the Affordable Care Act on Arkansas,} \emph{RAND Health Quarterly}, 3.

\bibitem[\protect\citeauthoryear{Sacarny, Baicker, and Finkelstein}{Sacarny
  et~al.}{2022}]{sacarny2022out}
\textsc{Sacarny, A., K.~Baicker, and A.~Finkelstein} (2022): \enquote{Out of
  the woodwork: enrollment spillovers in the Oregon health insurance
  experiment,} \emph{American Economic Journal: Economic Policy}, 14, 273--295.

\bibitem[\protect\citeauthoryear{Schmidt, Shore-Sheppard, and Watson}{Schmidt
  et~al.}{2019}]{schmidt2019impact}
\textsc{Schmidt, L., L.~Shore-Sheppard, and T.~Watson} (2019): \enquote{The
  Impact of Expanding Public Health Insurance on Safety Net Program
  Participation: Evidence from the ACA Medicaid Expansion,} Tech. rep.,
  National Bureau of Economic Research.

\bibitem[\protect\citeauthoryear{Simpson}{Simpson}{2020}]{simpson2020implications}
\textsc{Simpson, M.} (2020): \enquote{The implications of Medicaid expansion in
  the remaining states: 2020 update,} \emph{Washington, DC: Urban Institute},
  500.

\bibitem[\protect\citeauthoryear{Sommers and Epstein}{Sommers and
  Epstein}{2011}]{sommers2011states}
\textsc{Sommers, B.~D. and A.~M. Epstein} (2011): \enquote{Why states are so
  miffed about Medicaid—economics, politics, and the “woodwork effect”,}
  \emph{New England Journal of Medicine}.

\bibitem[\protect\citeauthoryear{Sommers and Gruber}{Sommers and
  Gruber}{2017}]{sommers2017federal}
\textsc{Sommers, B.~D. and J.~Gruber} (2017): \enquote{Federal funding
  insulated state budgets from increased spending related to Medicaid
  expansion,} \emph{Health Affairs}, 36, 938--944.

\bibitem[\protect\citeauthoryear{Sonier, Boudreaux, and Blewett}{Sonier
  et~al.}{2013}]{sonier2013medicaid}
\textsc{Sonier, J., M.~H. Boudreaux, and L.~A. Blewett} (2013):
  \enquote{Medicaid ‘welcome-mat’effect of Affordable Care Act
  implementation could be substantial,} \emph{Health Affairs}, 32, 1319--1325.

\bibitem[\protect\citeauthoryear{Sun and Abraham}{Sun and
  Abraham}{2021}]{sun2021estimating}
\textsc{Sun, L. and S.~Abraham} (2021): \enquote{Estimating dynamic treatment
  effects in event studies with heterogeneous treatment effects,} \emph{Journal
  of Econometrics}, 225, 175--199.

\bibitem[\protect\citeauthoryear{Truffer, Klemm, Wolfe, Rennie, and
  Shuff}{Truffer et~al.}{2013}]{cms2013}
\textsc{Truffer, C., J.~Klemm, C.~Wolfe, K.~Rennie, and J.~Shuff} (2013):
  \enquote{2013 Actuarial Report on the Financial Outlook for Medicaid,} Tech.
  rep., Centers for Medicaid \& Medicaid Services, Office of the Actuary.

\bibitem[\protect\citeauthoryear{Truffer, Rennie, Wilson, and Eckstein}{Truffer
  et~al.}{2018}]{cms2018}
\textsc{Truffer, C., K.~Rennie, L.~Wilson, and E.~Eckstein} (2018):
  \enquote{2018 Actuarial Report on the Financial Outlook for Medicaid,} Tech.
  rep., Centers for Medicaid \& Medicaid Services, Office of the Actuary.

\bibitem[\protect\citeauthoryear{Yocom}{Yocom}{2020}]{gao2020}
\textsc{Yocom, C.} (2020): \enquote{Accuracy of Determinations and Efforts to
  Recoup Federal Funds Due to Errors,} Tech. rep., Government Accountability
  Office, GAO-20-157.

\bibitem[\protect\citeauthoryear{Zhang and Zhu}{Zhang and
  Zhu}{2021}]{zhang2021does}
\textsc{Zhang, P. and L.~Zhu} (2021): \enquote{Does the ACA Medicaid Expansion
  Affect Hospitals’ Financial Performance?} \emph{Public Finance Review}, 49,
  779--814.

\end{thebibliography}

\section{Appendix} 
\begin{table}[H]
\renewcommand{\thetable}{A1}
\centering
\caption{State Medicaid Expansion Status}
\label{expansion_timing_table}
\footnotesize 
\renewcommand{\arraystretch}{0.8} 
\small
\begin{tabular}{@{} p{3.25cm} p{3.25cm} p{4.5cm} @{}}
\hline \hline
& & \\[-1.25ex]
State                & Implementation date & Designation in our analysis    \\ 
\toprule 
Alabama              & Has not expanded                     & Non-expansion  \\

Florida              & Has not expanded                     & Non-expansion  \\

Georgia              & Has not expanded                     & Non-expansion  \\

Kansas               & Has not expanded                     & Non-expansion  \\

Mississippi          & Has not expanded                     & Non-expansion  \\

South Carolina       & Has not expanded                     & Non-expansion  \\

Tennessee            & Has not expanded                     & Non-expansion  \\

Texas                & Has not expanded                     & Non-expansion  \\

Wisconsin            & Has not expanded                     & Non-expansion  \\

Wyoming              & Has not expanded                     & Non-expansion  \\

Arizona              & 1/1/2014                             & Expansion      \\

Arkansas             & 1/1/2014                             & Expansion      \\

California           & 1/1/2014                             & Expansion      \\

Colorado             & 1/1/2014                             & Expansion      \\

Connecticut          & 1/1/2014                             & Expansion      \\

Delaware             & 1/1/2014                             & Expansion      \\

District of Columbia & 1/1/2014                             & Expansion      \\

Hawaii               & 1/1/2014                             & Expansion      \\

Illinois             & 1/1/2014                             & Expansion      \\

Iowa                 & 1/1/2014                             & Expansion      \\

Kentucky             & 1/1/2014                             & Expansion      \\

Maryland             & 1/1/2014                             & Expansion      \\

Massachusetts        & 1/1/2014                             & Expansion      \\

Minnesota            & 1/1/2014                             & Expansion      \\

Nevada               & 1/1/2014                             & Expansion      \\

New Jersey           & 1/1/2014                             & Expansion      \\

New Mexico           & 1/1/2014                             & Expansion      \\

New York             & 1/1/2014                             & Expansion      \\

North Dakota         & 1/1/2014                             & Expansion      \\

Ohio                 & 1/1/2014                             & Expansion      \\

Oregon               & 1/1/2014                             & Expansion      \\

Rhode Island         & 1/1/2014                             & Expansion      \\

Vermont              & 1/1/2014                             & Expansion      \\

Washington           & 1/1/2014                             & Expansion      \\

West Virginia        & 1/1/2014                             & Expansion      \\

Michigan             & 4/1/2014                             & Expansion      \\

New Hampshire        & 8/15/2014                            & Expansion      \\

Pennsylvania         & 1/1/2015                             & Expansion      \\

Indiana              & 2/1/2015                             & Expansion      \\

Alaska               & 9/1/2015                             & Expansion      \\

Montana              & 1/1/2016                             & Expansion      \\

Louisiana            & 7/1/2016                             & Expansion      \\

Virginia             & 1/1/2019                             & Expansion      \\

Maine                & 1/10/2019                            & Expansion      \\

Idaho                & 1/1/2020                             & Non-expansion  \\

Utah                 & 1/1/2020                             & Non-expansion  \\

Nebraska             & 10/1/2020                            & Non-expansion  \\

Oklahoma             & 7/1/2021                             & Non-expansion  \\

Missouri             & 10/1/2021                            & Non-expansion  \\

South Dakota         & 7/1/2023                             & Non-expansion  \\

North Carolina       & 12/1/2023                            & Non-expansion  \\
\bottomrule
\end{tabular}
\caption*{{\footnotesize Source: Kaiser Family Foundation}} 
\end{table}



\clearpage
\vspace*{0.18\paperheight}

\begin{table}[H]
\centering
\renewcommand{\thetable}{A2}
\caption{Effects of Medicaid Expansion on Enrollment: Weighted by Medicaid Population}
\label{tab:weighted}
\renewcommand{\arraystretch}{1.2}
\begin{threeparttable}
\scriptsize
\begin{tabular}{l c c c c c c c c}
\toprule
 & (1) & (2) & (3) & (4) & (5) & (6) & (7) & (8) \\
\midrule
ATT & -0.0819\textsuperscript{**} & -0.1195\textsuperscript{***} & -0.0956\textsuperscript{**} & -0.0986\textsuperscript{**} & -0.1195\textsuperscript{***} & -0.1188\textsuperscript{***} & -0.1079\textsuperscript{**} & -0.0742\textsuperscript{*} \\
 & (0.0345) & (0.0392) & (0.0383) & (0.0405) & (0.0377) & (0.0445) & (0.0547) & (0.0394)\\
\addlinespace \midrule
Gov's Political Party & & \checkmark & \checkmark & \checkmark & \checkmark & \checkmark & \checkmark & \checkmark \\
Elig. Limit, Parents & & \checkmark & & \checkmark & \checkmark & \checkmark & \checkmark & \checkmark \\
Unemployment Rate & & \checkmark & \checkmark & & \checkmark & \checkmark & \checkmark & \\
Elig. Limit, Children & & & \checkmark & & & & & \\
Poverty Rate & & & & \checkmark & & & & \\
ln(State Population) & & & & & \checkmark & & & \\
Non-White (\% of State) & & & & & & \checkmark & & \\
TANF Benefits & & & & & & & \checkmark & \\
Food Insecurity Rate & & & & & & & & \checkmark \\
\addlinespace \midrule
N & 714 & 714 & 714 & 714 & 714 & 714 & 714 & 714 \\
\bottomrule
& & & & & & & & \\[-5.25ex]
\end{tabular}
\begin{tablenotes}
\scriptsize
    \item \textit{Notes}: This table shows estimates of the ATT of Medicaid expansion on the size of the original Medicaid population (logged) across a range of models, all of which use the difference-in-differences estimator described by \citet{callaway2021difference} weighted by each state's 2013 Medicaid population. The comparison group is "never" treated units. All other models include all states. Standard errors (clustered by state) are reported in parentheses. \\* $p<0.10$, ** $p<0.05$, *** $p<0.01$.
\end{tablenotes}
\end{threeparttable}
\end{table}

\begin{landscape}
\thispagestyle{landscapemode}

\vspace*{2cm}

\begin{table}[H]
\centering
\renewcommand{\thetable}{A3}
\caption{Effects of Medicaid Expansion on Enrollment in the Original Medicaid Population: Robustness of Main Specification}
\label{tab:robustnesspreferred}
\renewcommand{\arraystretch}{1.2}
\begin{threeparttable}
\scriptsize
\begin{tabular}{p{2.175cm} c c c c c c c c c c}
\toprule
 & (1) & (2) & (3) & (4) & (5) & (6) & (7) & (8) & (9) & (10)\\
\midrule
ATT (S.E.) & -0.0993\textsuperscript{***} & -0.1195\textsuperscript{***} & -0.1118\textsuperscript{***} & -0.1246\textsuperscript{***} & -0.0800\textsuperscript{***} & -0.1266\textsuperscript{**} & -0.0902\textsuperscript{**} & -0.1576\textsuperscript{***} & -0.1079\textsuperscript{***} & -0.1798\textsuperscript{***} \\
& (0.0348) & (0.0392) & (0.0395) & (0.0433) & (0.0350) & (0.0492) & (.0349) & (0.0578) & (0.0414) & (0.0684) \\
\midrule
Gov Pol Party & \checkmark & \checkmark & \checkmark & \checkmark & \checkmark & \checkmark & \checkmark & \checkmark & \checkmark & \checkmark \\
Elig Parents  & \checkmark & \checkmark & \checkmark & \checkmark & \checkmark & \checkmark & \checkmark & \checkmark & \checkmark & \checkmark \\
Unemployment  & \checkmark & \checkmark & \checkmark & \checkmark & \checkmark & \checkmark & \checkmark & \checkmark & \checkmark & \checkmark \\
\midrule
Drop Late Exp &  &  & \checkmark & \checkmark &  &  &  &  & \checkmark & \checkmark \\
Drop 2010-12 Exp &  &  & & & \checkmark & \checkmark &  &  & \checkmark & \checkmark \\
Drop Pre-2010 Exp &  &  & & & & & \checkmark & \checkmark & \checkmark & \checkmark \\
\midrule
Weighted & & \checkmark & & \checkmark & & \checkmark & & \checkmark & & \checkmark \\
\bottomrule
N & 714 & 714 & 602 & 602 & 630 & 630 & 574 & 574 & 462 & 462 \\
\bottomrule
& & & & & & & & & \\[-5.25ex] 
\end{tabular}
\begin{tablenotes} 
    \item \textit{Notes}: This table shows estimates of the ATT of Medicaid expansion on the size of the original Medicaid population (logged) across a range of specifications, all of which use the difference-in-differences estimator described by \citet{callaway2021difference}. Specifications that drop Late Expanders exclude the following states from the sample: AK, IN, LA, ME, MT, NH, PA, and VA. Specifications that drop 2010-2012 Expanders exclude the following states: CA, CT, DC, MN, NJ, WA. Specifications that drop Pre-2010 Expanders exclude the following states: DE, DC, MA, NY, VT. The comparison group is ``never-treated" units. All other models include all states. Standard errors (clustered by state) are reported in parentheses. \\ * $p<0.10$, ** $p<0.05$, *** $p<0.01$.
\end{tablenotes}
\end{threeparttable}

\end{table}

\end{landscape}

\pagebreak 

\vspace*{0.15\paperheight}

\begin{table}[H]
\centering
\renewcommand{\thetable}{A4}
\caption{Effects of Medicaid Expansion on Enrollment Original Medicaid Population: 2014 Cohort Only}
\label{tab:early_expanders}
\renewcommand{\arraystretch}{1.2}
\begin{threeparttable}
\scriptsize
\begin{tabular}{l c c c c c c c c}
\toprule
 & (1) & (2) & (3) & (4) & (5) & (6) & (7) & (8) \\
\midrule
ATT (S.E.) & -0.0595 & -0.1118\textsuperscript{***} & -0.0777\textsuperscript{***} & -0.1209\textsuperscript{***} & -0.1040\textsuperscript{**} & -0.1021\textsuperscript{**} & -0.1159\textsuperscript{**} & -0.0755\textsuperscript{**} \\
 & (0.0370) & (0.0395) & (0.0362) & (0.0417) & (0.0382) & (0.0429) & (0.0496) & (0.0353) \\ \midrule
\addlinespace
Gov's Political Party & & \checkmark & \checkmark & \checkmark & \checkmark & \checkmark & \checkmark & \checkmark \\
Elig. Limit, Parents & & \checkmark & & \checkmark & \checkmark & \checkmark & \checkmark & \checkmark \\
Unemployment Rate & & \checkmark & \checkmark & & \checkmark & \checkmark & \checkmark & \\
Elig. Limit, Children & & & \checkmark & & & & & \\
Poverty Rate & & & & \checkmark & & & & \\
ln(State Population) & & & & & \checkmark & & & \\
Non-White (\% of State) & & & & & & \checkmark & & \\
TANF Benefits & & & & & & & \checkmark & \\
Food Insecurity Rate & & & & & & & & \checkmark \\
\addlinespace \midrule
N & 602 & 602 & 602 & 602 & 602 & 602 & 602 & 602 \\
\bottomrule
& & & & & & & & \\[-5.25ex]
\end{tabular}
\begin{tablenotes}
\scriptsize
    \item \textit{Notes}: This table shows estimates of the ATT of Medicaid expansion on the size of the original Medicaid population (logged) across a range of models, all of which use the difference-in-differences estimator described by \citet{callaway2021difference}. The treatment group in this table is states that expanded in 2014. The comparison group is ``never-treated" units. All other models include all states. Standard errors (clustered by state) are reported in parentheses. \\* $p<0.10$, ** $p<0.05$, *** $p<0.01$.
\end{tablenotes}
\end{threeparttable}
\end{table}

\setcounter{figure}{0}
\renewcommand{\thefigure}{A\arabic{figure}}
\vspace*{3cm}

\begin{figure}[H]
\centering

\caption{\label{appendix_results_plot_weighted} Dynamic Effects of Medicaid Expansions on Enrollment: Weighted by Medicaid Population}
\includegraphics[width=\textwidth]{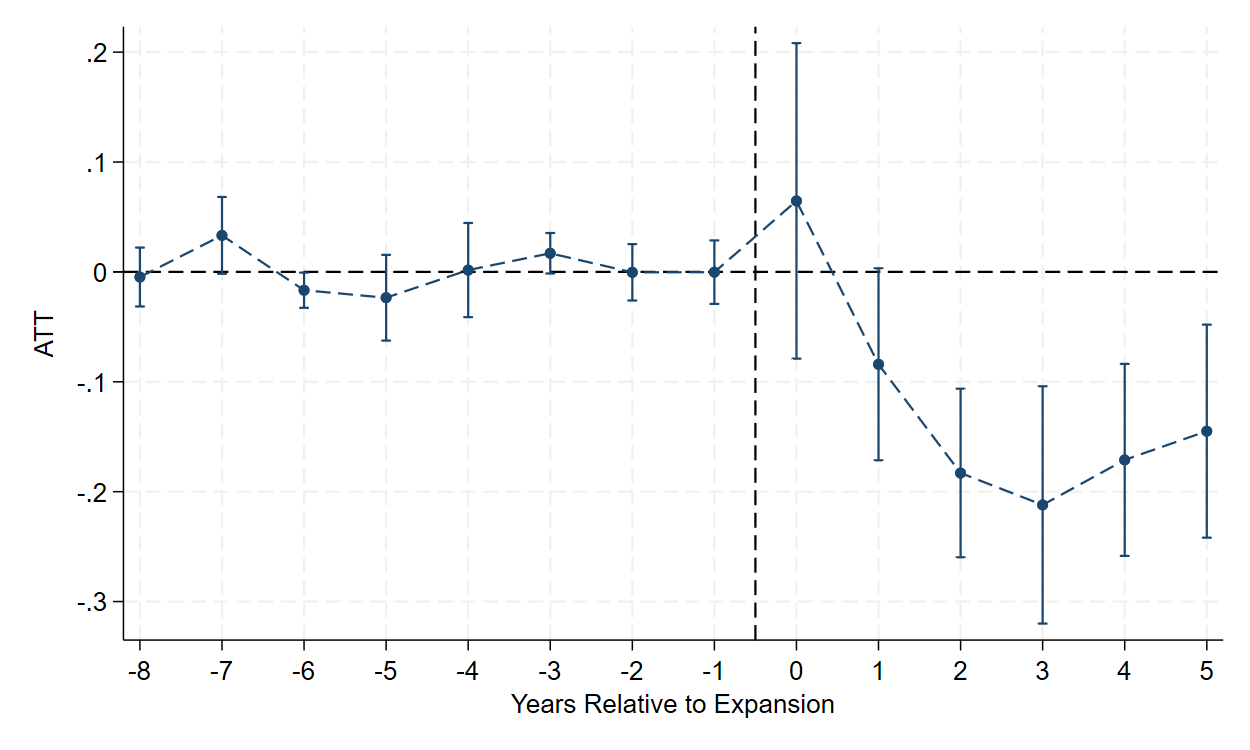}
\caption*{{\footnotesize \textit{Notes}: This plot shows dynamic effects across event time, from (column (3) in Table \ref{tab:weighted}), which are weighted by each state's 2013 Medicaid population. Bars represent 95\% confidence intervals. The vertical dashed line represents the implementation of Medicaid expansion. We use enrollment figures for the month of June in each year. Data was compiled by the authors from Kaiser Family Foundation issue briefs (for the years 2006-2013) and reports from the Medicaid Budget and Expenditure System (for the years 2014-2019); see Section \ref{Data} for more details. We define the original Medicaid population as total Medicaid enrollment minus the number of enrollees reported by states as ``newly eligible" under the ACA.} }
\end{figure}

\setcounter{figure}{1}
\renewcommand{\thefigure}{A\arabic{figure}}
\vspace*{3cm}

\begin{figure}[H]
\centering

\caption{\label{appendix_2014_states_results_plot} Dynamic Effects of Medicaid Expansion on Enrollment: 2014 Cohort Only}
\includegraphics[width=\textwidth]{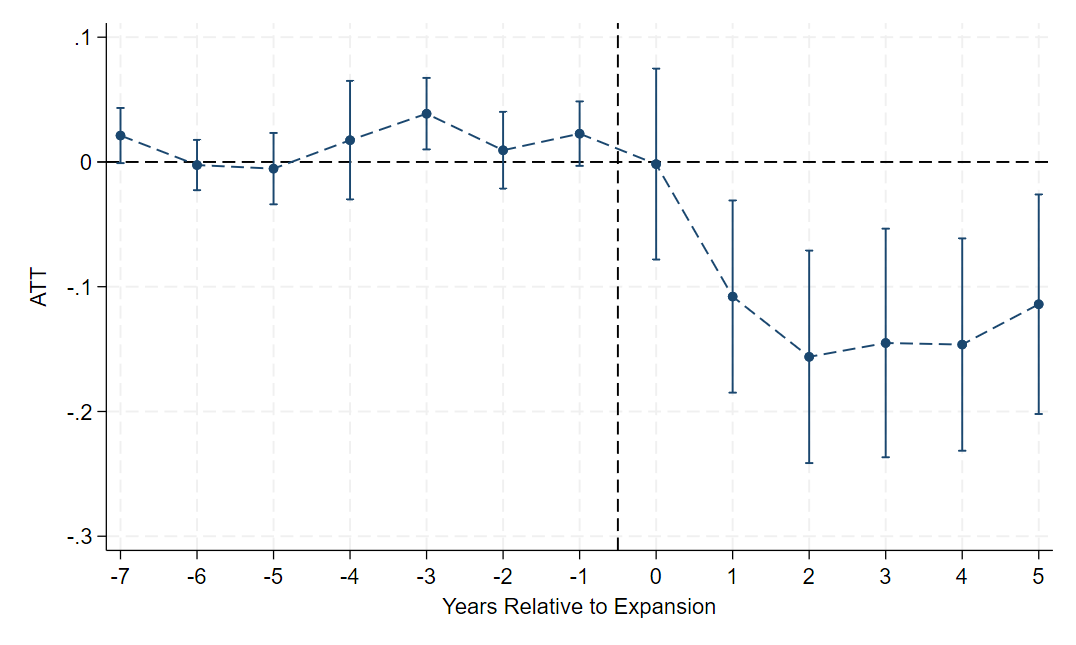}
\caption*{{\footnotesize \textit{Notes}: This plot shows dynamic effects across event time, from (column (3) in Table \ref{tab:early_expanders}). Bars represent 95\% confidence intervals. The vertical dashed line represents the implementation of Medicaid expansion. We use enrollment figures for the month of June in each year. Data was compiled by the authors from Kaiser Family Foundation issue briefs (for the years 2006-2013) and reports from the Medicaid Budget and Expenditure System (for the years 2014-2019); see Section \ref{Data} for more details. We define the original Medicaid population as total Medicaid enrollment minus the number of enrollees reported by states as ``newly eligible" under the ACA.} }
\end{figure}

\vspace*{3cm}

\setcounter{figure}{2}
\vspace{.5cm}
\begin{figure}[H]
\centering
\caption{\label{data_check_scatterplot} Medicaid Enrollment in 2013 by Data Source}
\includegraphics[width=.9\textwidth]{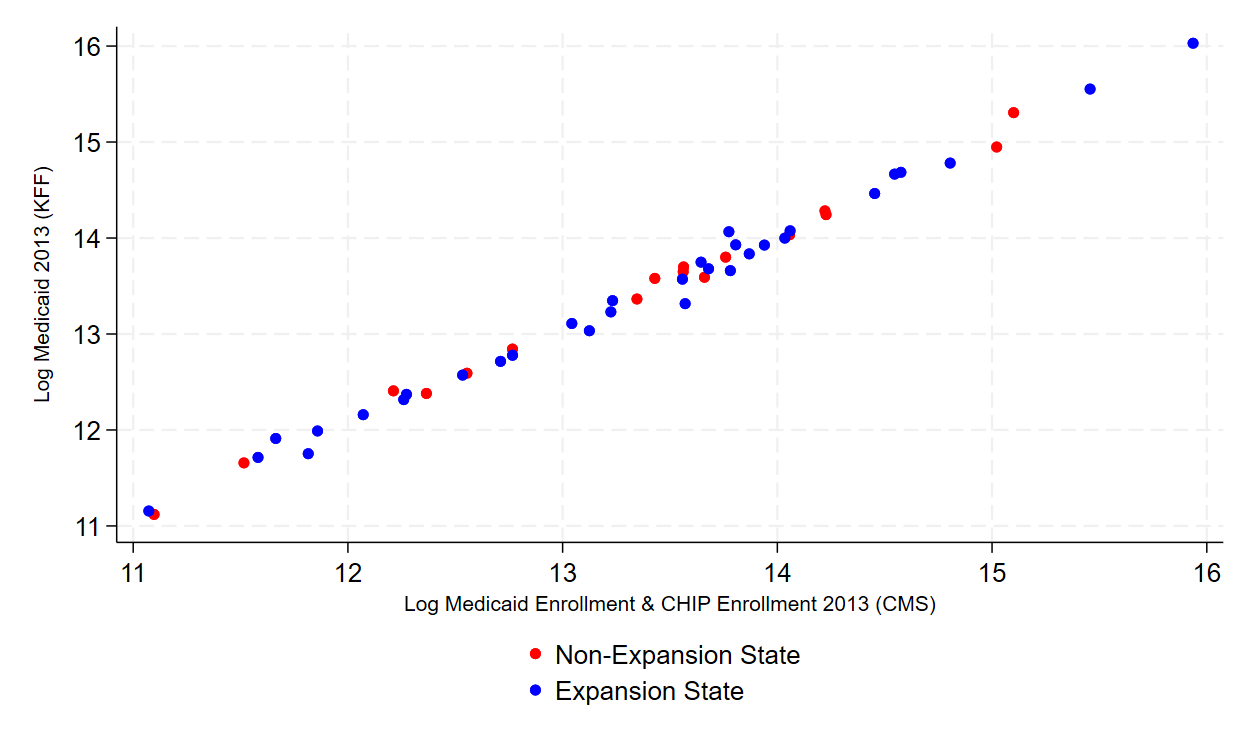}
\caption*{{\footnotesize \textit{Notes}: This figure plots the (log) Total Medicaid Population for December 2013 from CMS-MBES against the (log) Total Medicaid and CHIP population in KFF reports from 2013. This graph excludes Connecticut and Maine as data was not available for these states for December 2013 in CMS-MBES. Data was compiled by the authors from Kaiser Family Foundation issue briefs (for the years 2006-2013) and reports from the Medicaid Budget and Expenditure System (for the years 2014-2019); see section \ref{Data} for more details.} }
\end{figure}

\end{document}